\documentclass[twocolumn,showpacs,eqsecnum,amsmath,amssymb,aps,prb,floatfix]{revtex4-1}

\usepackage{graphicx} 
\usepackage{dcolumn} 
\usepackage{bm} 
\usepackage{hyperref} 
\usepackage{slashed}
\usepackage{subfigure}

\begin{document}

\title{Field Theory of Nematicity in the Spontaneous Quantum Anomalous Hall effect}

\author{Yizhi You}
\author{Eduardo Fradkin}
\affiliation{
Department of Physics and Institute for Condensed Matter Theory, University of Illinois at Urbana-Champaign, 1110 West Green Street, Urbana, Illinois, 61801-3080
}

\date{\today}

\begin{abstract}
We derive from a microscopic model the effective theory of nematic order in a system 
with a spontaneous quantum anomalous Hall effect in two dimensions. 
Starting with a model of two-component fermions (a spinor field) with a quadratic band crossing and short range 
four-fermion marginally relevant interactions we use a $1/N$ expansion and bosonization methods to derive the 
effective field theory for the hydrodynamic modes associated with the conserved currents and with the local fluctuations 
of the nematic order parameter. We focus on the vicinity of the quantum phase transition from the isotropic 
Mott Chern insulating phase to a phase in which time-reversal symmetry breaking coexists with nematic order, 
the nematic Chern insulator. The topological sector of the effective field theory is a BF/Chern-Simons gauge theory. 
We show that the nematic order parameter field couples with the Maxwell-type terms of the gauge fields as the space components 
of a locally fluctuating metric tensor. 
The nematic field has $z=2$ dynamic scaling exponent.  The low-energy dynamics  of the nematic order parameter is found to be 
governed by a Berry phase term. By means of a detailed analysis of the coupling of the spinor field of the fermions to the changes 
of their local frames originating from long-wavelength lattice deformations we calculate the Hall viscosity of this system and show 
that in this system it is not the same as the Berry phase term in the effective action of the nematic field, but both are related to the concept of torque Hall viscosity which we introduce here. 
\end{abstract}
\maketitle

\section{Introduction and motivation}
\label{sec:introduction}

The theory of topological phases of matter has been a central problem in condensed matter physics since the discovery of the 
quantum Hall effects\cite{Klitzing-1980,Tsui-1982} in two-dimensional electron gases (2DEG) in large magnetic fields. 
The precisely observed (quantized or fractional) 
values of the Hall conductance is a manifestation of the fact that it is a topological invariant of 
the incompressible fluid.\cite{Laughlin-1981,Thouless-1982,Niu-1985} 
The fractional quantum Hall fluids, on the other hand, are explained by the universal properties  
encoded in the structure of  their wave functions\cite{Laughlin-1983} whose excitations (vortices)
carry fractional charge and fractional statistics.\cite{Laughlin-1983,Haldane-1983,Halperin-1984}  
The robustness of these properties a consequence of their topological character.
In addition to  having fractionalized excitations, these topological fluids 
have a ground state degeneracy which depends on the topology of the surface on which they reside,
which is not a consequence of the spontaneous breaking of any global symmetry.\cite{Wen-1990} 
The universal behavior of these topological fluids is encoded in an effective low-energy, the
 Chern-Simons gauge theory.\cite{Zhang-1989,Lopez-1991,Frohlich1991,Wen-1992,Wen-1995}

There is now a growing body of (mostly theoretical) evidence that such topological phases of matter  
exist in several models of 
frustrated quantum antiferromagnets\cite{Jiang-2011}   and
in quantum dimer models.\cite{Rokhsar-1988,Moessner-2001} 
The recent discovery of topological insulators\cite{Bernevig-2006b,Konig-2007,Fu-2007b,Hasan-2010,Hasan-2011} 
has opened a new arena in 
which these ideas play out. Interacting versions of simple models of topological Chern insulators, such as the Haldane model,\cite{qah} 
have topological phases with fractionalized excitations.\cite{Neupert-2011,Sheng-2011,Regnault-2011,Cinicio-2013}

An interesting  question is the interlay and possible coexistence of  topological order 
and spontaneous symmetry breaking.
For some filling fraction the 2DEG is known to have a ferromagnetic quantum Hall ground state,\cite{Sondhi-1993,Ho-1994} 
in which  spin rotational symmetry is spontaneously broken. 
Also, a state with a nematic ``valley'' order has also been seen in quantum Hall fluids 
on misoriented samples.\cite{Shkolnikov-2005,Abanin-2010}
 On the other hand, experiments in the 2DEG in the second Landau level found a nematic state in a regime in which the 
 fractional (and integer) quantum Hall effect is absent.\cite{Lilly-1999,Pan-1999,Cooper-2002} In this phase the 2DEG is
an uniform gapless electron fluid with a spontaneously broken spatial rotational symmetry.\cite{Fradkin-1999,Fradkin-2000} 
 
Recent experiments by Xia and coworkers  found that the 2DEG in the first Landau level in tilted 
magnetic fields has a strong 
 tendency to break rotational invariance inside an incompressible
 fractional quantum Hall Laughlin state.\cite{Xia-2010,Xia-2011}
 Although in the the experiments rotational invariance is broken explicitly by the tilted magnetic field, the temperature dependence of 
 the transport anisotropy suggest that this state has a large nematic susceptibility and may be close to a phase transition to a nematic state. 
 These experiments  motivated Mulligan, Kachru and Nayak to develop a theory in 
 which nematic order coexists with a fractional quantum Hall 
 fluid.\cite{chetan,Mulligan-2011}  The possible existence of such  states was anticipated by two early proposals of wave 
 functions for anisotropic quantum Hall fluids.\cite{Balents-1996,Musaelian-1996} 
 
 The experiments of Xia and coworkers have also 
 motivated the  inquiry  of the role of more microscopic,  ``geometrical'', degrees of freedom in the physics of these 
 topological fluids.\cite{Haldane-2011,haldanemetric,Qiu-2012} 
 Recently, Maciejko and coworkers proposed an effective field theory of the anisotropic 
 fractional quantum Hall state.\cite{Maciejko-2013} Using mainly  symmetry arguments, they found that the nematic order parameter 
 couples to the fractional quantum Hall fluid in the same way as the space components of a metric tensor. 
 A similar effect was found earlier in a theory of a nematic charge $4e$ superconductor\cite{Barci-2011}  involving, instead, the order 
 parameter field of the superconductor.
 A key result of Ref.[\onlinecite{Maciejko-2013}] is that the dynamics of the nematic degrees of freedom is 
 governed by a Berry phase term in the effective action whose coefficient is the Hall 
 viscosity of the topological fluid.\cite{avron1995,Read-2009,taylor,Nicolis-2011,vis}

There are many aspects of this problem which remain unclear. In the case of the 2DEG the  existence of a compressible nematic phase (in 
the second Landau level) suggests that it must be related to the anisotropy seen in the first Landau level, albeit in the incompressible phase. 
The theory of Ref.[\onlinecite{chetan}] suggests a possible mechanism (and an identification of the nematic degrees of freedom) solely in 
terms of the low-energy degrees of freedom of the quantum Hall fluid, but runs into difficulties in systems with Galilean invariance. In addition, 
that theory should also apply to the case of the integer quantum Hall effect. Although it is possible to write down a wave function for an 
anisotropic quantum Hall state by breaking rotational invariance explicitly at the microscopic level,\cite{haldanemetric} such an approach 
does not explain how it may come about from an isotropic incompressible state. 

In this paper we will investigate these problems by deriving an effective field theory for a Mott Chern insulator in a nematic phase in a
 a simple microscopic lattice model recently proposed in by Sun and coworkers.\cite{kai} 
We will discuss in detail the case of the 2DEG in magnetic fields in a  separate publication.\cite{You-2013}
The  model of Ref.[\onlinecite{kai}] describes a correlated two-dimensional system of spinless fermions on a checkerboard square lattice in which 
two bands have a 
quadratic crossing at the corners of the (square) Brillouin zone. 
In the non-interacting system the quadratic band crossing is protected by the $C_4$ point group symmetry of square lattice and 
by time-reversal invariance.

Due to the quadratic band crossing, this electronic system has a dynamical scaling exponent $z=2$ (i.e. the 
energy scales with the square of the momentum). 
As a direct consequence of the $z=2$ scaling, four fermion operators are naively marginal operators.
This free-fermion system, which can be regarded 
as a fermionic version of a quantum Lifshitz model,\cite{Ardonne2004} is at an infrared unstable fixed point of the renormalization group (RG).  
This semimetal fixed point
is unstable to infinitesimal repulsive interactions to a) a gapped phase with a spontaneously broker time-reversal invariance,
i.e. a topological Mott Chern insulator with a spontaneous quantum anomalous Hall state\cite{qi}, b) to
a gapless semimetal nematic phase in which the point group symmetry breaks spontaneously from $C_4$ down to $C_2$, 
and c) to a gapped phase in which both time reversal symmetry-breaking and the point group symmetry breaking coexist.\cite{kai} 

Models with quadratic band crossings  describe the low-energy description of graphene bilayers,\cite{Nandkishore2010,Vafek2010,Lemonik2012} where there are two such crossings, and in the topologically protected surface states of 3D topological crystalline insulators.\cite{Fu2011,Fang2012} We discuss below some caveats on the relevance of this model to such systems. In particular, a (Mott) Chern insulating state has been conjectured to exist in bilayer graphene.\cite{Nandkishore2010}

Due to the  marginal relevance of local interactions, the behavior of the system in these phases 
can be investigated using controlled approximations, such as $1/N$ expansions and perturbative RG calculations.
In contrast, in the case of the massless Dirac fermion, local interactions 
are irrelevant and a finite (and typically large) critical value of the coupling constant is need to drive the system into a 
Mott Chern insulating phase.\cite{qi} Here we will use the $1/N$ expansion and bosonization methods to derive an effective field 
theory of the Mott Chern insulator and of its quantum phase transition to a nematic Chern insulator in the context of the model of 
Ref.[\onlinecite{kai}]. The effective field theory includes a the hydrodynamic degrees of freedom of the conserved currents of the 
fermions, in the form of a BF/Chern-Simons gauge theory,
and the local fluctuations of the nematic order parameter. In particular in this theory the nematic fluctuations are present at low 
energies which is required to describe a continuous quantum phase transition to a nematic Mott Chern insulator. 

We will also show that the effective low-energy dynamics of the nematic order parameter is indeed a Berry phase term, 
with a structure similar to that proposed by Maciejko and collaborators. 
We  also find that the nematic fields  can be regarded as providing a local fluctuating spatial metric 
for the hydrodynamic gauge fields of the Mott Chern insulator. 
However we will also show that the nematic degrees of freedom do not couple to the fermionic degrees of freedom 
as a local frame field and hence, they cannot be identified with a local geometry. 
We show that the Hall viscosity, which in system of spinors is the response of the system to a  change of the local frames\cite{taylor} 
(i.e. a long-wavelength distortion of the   lattice), is not equal to the Berry phase of the nematic modes. 
Instead,  the Berry phase is related to the concept of torque Hall viscosity which we introduce here. 
In addition, we find that in this system the Hall viscosity is not given by the coefficient of the  $q^2$ term in the Hall conductance. 
Recently, Hoyos and Son showed that in Galilean-invariant one-component quantum Hall fluids systems these two coefficients should be equal to each other.\cite{vis} These assumptions do not apply to multi-component fermionic (spinor) systems as in the present case.
We also find that the Hall viscosity and the Berry phase coefficient are related to the  Hall torque viscosity.

This paper is organized as follows. In Section \ref{sec:model} we present the model of interacting fermions in two dimensions 
with a quadratic band crossing and we discuss its phase diagram. In Section \ref{sec:effective-field-theory} we develop an 
effective field theory of the interplay of nematic order and of the hydrodynamic gauge theory. 
In Section \ref{sec:1/N} we use the $1/N$ expansion to derive the effective action in the vicinity of the nematic transition 
inside the spontaneous quantum anomalous Hall phase, and use it to discuss briefly the nature of the two phases and the quantum and thermal critical behavior. In Section \ref{sec:nematic-effective-action} we present the effective field 
theory of the nematic fields in the presence of broken time-reversal invariance. 
Here we discuss in detail the role played by the Hall viscosity in the effective field theory. 
In Section \ref{sec:torque-viscosity} we introduce the concept of Hall torque viscosity and discuss its relation with the Hall viscosity and with the Berry phase. 
Our conclusions are presented in Section \ref{sec:conclusions}. Details of the calculations, including the  proofs of gauge invariance, 
are given in several appendices.

\section{The Quadratic Band-Crossing Model and its Phases}
\label{sec:model}

In this paper we will use the following simple model for a quadratic band crossing (QBC),introduced by Sun and collaborators.\cite{kai} 
We begin with a summary of the results of their work that will be useful for our analysis. On of the cases discussed by Sun {\it et. al} is 
a system of spinless fermions on a checkerboard lattice. This lattice has two sublattices, and the single-particle states are two-component spinors.
The band structure of this system is described by the tight-binding one-particle Hamiltonian
\begin{align}
h_{0}(\textbf{k})=&t(\cos k_1-\cos k_2)\sigma_3+4t' \cos\left(\frac{k_1}{2}\right)\cos\left(\frac{k_2}{2}\right)\sigma_1
\nonumber\\
&
\label{eq:h0-QBC}
 \end{align}
 where $\textbf{k}=(k_1,k_2)$ are vectors of the first Brillouin zone (BZ), $|k_i|\leq \pi$ (with $i=1,2$), 
 and $\sigma_1$ and $\sigma_3$ are two (real symmetric, $2 \times 2$) Pauli matrices. 
The lattice model also has a contribution proportional to the $2\times 2$ identity matrix which,
for a range of parameters, can be ignored.\cite{sun2011} Tsai and coworkers\cite{tsai-2011} discussed a similar problem on the Lieb lattice.

 In this system the two bands cross at the Fermi energy at the corners of the BZ, $(\pi,\pi)$ (and its symmetry related points). 
 For a half-filled system, the Fermi energy is exactly at the band crossing points, and the ground state of the non-interacting system describes a semi-metal with a 
 quadratic band dispersion. Similar problems have been discussed in the context of bilayer graphene.\cite{Nandkishore2010,Vafek2010,Vafek2010b}
 
 The band structure of this semi-metal has a non-trivial Berry phase
\begin{eqnarray}
i \oint_\Gamma d\textbf{k}\cdot \langle \textbf{k}|\nabla_{\textbf{k}}|\textbf{k}\rangle=n\pi 
 \end{eqnarray}
where $|\textbf{k}\rangle$ is a Bloch state at momentum $\textbf{k}$ of the BZ, and $\Gamma$ is a closed curve on the 
BZ that encloses the quadratic band crossing point, $(\pi,\pi)$. For a two-band system with a QBC the integer $n=2$ ($n=\pm 1$ for Dirac fermions). 
In this case the changes of the Chern number of the two bands are carried entirely by the (single) quadratic crossing. 
At the non-interacting level, the Berry phase here is protected by both discrete lattice symmetries and by time reversal invariance. 
 
 For momenta $\textbf{k}=(\pi,\pi)-\textbf{q}$ close to the crossing points (the corners of the BZ) we can approximate the one-particle 
 Hamiltonian by expanding Eq.\eqref{eq:h0-QBC} about the crossing point. Let us denote by $\psi_\alpha(\textbf{q})$ (with $\alpha=1,2$) 
 be a two-component Fermi field with wave vectors $\textbf{q}$ (measured from the $(\pi,\pi)$ point). 
 The effective free fermion Hamiltonian, in momentum and in position space, is
\begin{align}
H_0=&\int \frac{d^2q}{(2\pi)^2} \psi^\dagger_\alpha(\textbf{q})  \Big( (q_{1}^2-q_{2}^2) \sigma_3 +2 q_1 q_2 \sigma_1 \Big)_{\alpha \beta} \psi_\beta(\textbf{q})\nonumber\\
      =-&\int d^2x \; \psi^\dagger_\alpha(\textbf{x})  \Big(\sigma_3 \; (\partial_1^2-\partial_2^2) + \sigma_1 \; 2\partial_1 \partial_2  \Big)_{\alpha \beta} \psi_\beta(\textbf{x})
\label{eq:H0-long-wavelengths}
 \end{align}
Here, and from now on,  we have set $t=t'$ for simplicity (and rescaled the energy scale so that $t=1$). 
This is a special point of high (rotational) symmetry which
does not  qualitatively change the results. In the case of bilayer graphene one has two ``valleys'' (or species) of fermions whose free-fermion Hamiltonians are given by Eq.\eqref{eq:H0-long-wavelengths}, except that the sign of $t'$, a chirality that distinguishes one valley from the other. Thus for bilayer graphene one has $|t|=|t'|$.

For a system of (spinless) fermions with a QBC with short-range repulsive microscopic interactions, the effective low-energy Hamiltonian is the sum of the free-fermion Hamiltonian  $H_0$ of Eq.\eqref{eq:H0-long-wavelengths} and an interaction term $H_{\rm int}$ which can be succinctly written in the form
\begin{equation}
H_{\rm int}=-\int d^2x \; \frac{1}{2} \Big(g_0 \Phi_0^2(\textbf{x})+g {\bm \Phi}^2(\textbf{x})\Big)
\label{eq:Hint-spinless}
\end{equation}
where $g_0$ and $g$ are two (positive) coupling constants. The operators $\Phi_0(\bm{x})$ and ${\bm \Phi}(\bm{x})$ in Eq.\eqref{eq:Hint-spinless} are, respectively, given by 
the (Hermitian) bilinears of fermion operators, 
\begin{align}
\Phi_0({\bm x})=&\psi^\dagger(\textbf{x}) \sigma_2 \psi(\textbf{x})\label{eq:time-reversal-order-parameter}\\
{\bm \Phi}(\bm{x})=& \psi^\dagger(\textbf{x}) \bm{\sigma} \psi(\textbf{x})
\label{eq:nematic-order-parameter}
\end{align}
Here $\bm{\sigma}=(\sigma_1,\sigma_3)$, and, for clarity,  where we have suppressed the spinor indices. For $t=t'$ the  full Hamiltonian, $H=H_0+H_{\rm int}$ is invariant under time-reversal and under arbitrary rotations. 
However for $t \neq t'$, it is only invariant under the (discrete) point-group $C_{4}$. 

The operator $\Phi_0(\bm{x})$ of Eq.\eqref{eq:time-reversal-order-parameter} breaks time-reversal invariance and is the order parameter for time-reversal symmetry breaking.
If $\langle \Phi \rangle \neq 0$ the system would have a gap and exhibit a zero-field quantum Hall effect with $\sigma_{xy}=e^2/h$ (i.e. an anomalous quantum Hall effect).
 The operator ${\bm \Phi}(\bm{x})$ of Eq.\eqref{eq:nematic-order-parameter} breaks rotational invariance and  it is the  nematic order parameter. 
 In fact $\bm{\Phi}$ is invariant under a rotation by $\pi$ and hence it is not a vector but a director, as it should be. 
Moreover, if  we were to add terms proportional to the operators $\Phi_0$ and ${\bm \Phi}$ to the free-fermion Hamiltonian of Eq. \eqref{eq:H0-long-wavelengths}, 
  the QBC either gets gapped (if $\langle \Phi_0\rangle \neq 0$) or splits  into two massless Dirac fermions which are separated either along the $x$ (or $y$) axis (is $\langle \Phi_1 \rangle \neq 0)$) or along a diagonal (if $\langle \Phi_2 \rangle \neq 0$). Hence this state breaks  rotational invariance (or $C_4$ or $C_6$ down to $C_2$). 
Hence, a state with $\langle \Phi_0 \rangle \neq 0$ is a topological Chern insulator, 
while a state with $\langle {\bm \Phi}\rangle \neq 0$ is a nematic semi-metal. 
If spin and other degrees of freedom are also considered, other operators (and hence possible phases) which transform non-trivially under other  symmetries must be considered, leading, for instance, 
 to a state with a spin Hall effect, a ferromagnet, triplet nematic order, and others.\cite{kai,Nandkishore2010,Vafek2010,Vafek2010b,Lemonik2012}

In the case of  the theory of a massless Dirac fermions (e.g., graphene) short-range interactions are  irrelevant operators, rendering the semi metallic phase stable, 
and can only trigger a (quantum) phase transition if the coupling constants are larger than a critical value.\cite{wilson-1970} 
However, in the case of a theory of fermions with a QBC, short-range interactions of the form of Eq.\eqref{eq:Hint-spinless} are marginally relevant and destabilize the QBC semimetal even for arbitrarily weak interactions\cite{kai} (see also the prescient work of Abrikosov and coworkers\cite{Abrikosov1974}).

The kinematic differences between the two systems, Dirac and the QBC,  leads to a change in the scaling behavior of the operators.\cite{kai} 
In particular the Hamiltonian $H_0$ of Eq.\eqref{eq:H0-long-wavelengths} 
describes a quantum critical system of free fermions with dynamical exponent $z=2$ and, hence, in this system time scales as the square of a length, $L^2$. 
For this reason it has some similarities with systems in the quantum Lifshitz universality class.\cite{Ardonne2004} 
Consequently, in a system with $z=2$ dynamic scaling, in two space dimensions the fermion operator has  scaling dimension $\Delta_\psi=1$, $[\psi]=L^{-1}$, and all four-fermion operators 
have scaling dimension $4$. 

In {\em two} (space) dimensions this means that all four fermion operators are {\em marginal} (in the renormalization group (RG) sense) since here $d+z=4$. Therefore, the stability (or instability) of the free-fermion QBC semimetal, such as the surface states of the three-dimensional crystalline topological insulators,\cite{Fu2011,Hasan2012,Yazdani2013} such as Pb$_{1-x}$Sn$_x$Te, is determined by quantum corrections.
In contrast, systems with a QBC in {\em three} dimensions, 
such as the pyrochlore iridates A$_2$Ir$_2$O$_7$ (where A is a lanthanide or yttrium) ,\cite{Yang2010,Witczak-Krempa2012,Witczak-Krempa2013} short-range interactions 
are perturbatively irrelevant and the QBC semimetal is stable (up to a critical value of the coupling constants) (see, once again, Ref.[\onlinecite{Abrikosov1974}]). 

One-loop renormalization group calculations show that, in two dimensions, in a system with microscopic repulsive interactions, and hence $g_0>0$ and $g>0$,
four-fermion operators of the form of Eq.\eqref{eq:Hint-spinless}
 are {\em marginally relevant},\cite{kai,Vafek2010,Vafek2010b} and, hence, weak repulsive interactions render the semi-metal free-fermion ground state unstable. 
Several phases can occur depending on the details of the microscopic interactions.
 In Ref.[\onlinecite{kai}] it was shown that in the case of the QBC of the checkerboard lattice a weak (infinitesimal) repulsive interaction drives the system into a state with a spontaneous anomalous quantum Hall effect (i.e. a Chern insulator with a spontaneously-broken time-reversal symmetry), with a subsequent phase transition to a nematic semimetal state. Sun and coworkers\cite{kai} also found a regime in which the nematic state and the  Chern insulating state coexist. Thus, in this phase, the system has a spontaneously broken time-reversal invariance and also a spontaneously broken rotational invariance, and is a nematic Chern insulator. Such topological Mott insulators were proposed earlier on by Raghu, Qi, Honerkamp and Zhang in the context of Dirac-type systems where they can only occur at relatively large values of the interactions.\cite{qi}

 \section{Effective gauge theory for the anisotropic QAH state}
 \label{sec:effective-field-theory}
 
Our goal is to derive an effective action for the spontaneous QAH phase and to describe the transition to a nematic QAH phase. 
To this end we will generalize our system to one in which there are $N$ ``flavors" of fermions and to drive the effective field theory using a large-$N$ expansion. 
Sun and coworkers have shown that, unlike the familiar case of the Luttinger liquids in one space dimensions, 
the renormalization group beta function(s) for the $N=1$ case has the same structure as the $N>1$ case.\cite{kai} 
The resulting effective Lagrangian density for the spinor fermionic field $\psi_{a}(x)$ (with $a=1,\ldots,N$, $x=(x_0,\vec x)$, and $x_0$ is the time coordinate) 
(here we are omitting the spinor indices)
\begin{widetext}
\begin{equation}
\mathcal{L}_F[\bar \psi,\psi,a_\mu]=\bar \psi_a (x) \Big(i \gamma_0 D_0-  \gamma_1(D_1^2-D_2^2)-  \gamma_2 (D_1 D_2+D_2D_1) \Big) \psi_a(x)+ \frac{g_0}{2N} \Phi_0(x)^2+\frac{g}{2N} {\bm \Phi}^2(x)
\label{eq:Lagrangian}
\end{equation}
\end{widetext}
where $\Phi_0(x)$ and ${\bm \Phi}(x)$ are the fermion bilinears defined in Eq.\eqref{eq:time-reversal-order-parameter} and Eq.\eqref{eq:nematic-order-parameter}, 
respectively, suitably generalized for a system with $N$ flavors of fermions. Minimal coupling of the fermions to the gauge field requires that we change of the Hamiltonian of the system to insure its hermiticity and gauge invariance.

In Eq.\eqref{eq:Lagrangian}
we have used the standard $2 \times 2$ Dirac gamma matrices, given in terms of the three Pauli matrices 
\begin{equation}
\gamma_0=\sigma_2, \qquad \gamma_1=i \sigma_1, \qquad  \gamma_2=-i \sigma_3
\end{equation}
and satisfy the Dirac (Clifford) algebra (with $\mu=0,1,2$)
\begin{equation}
\{ \gamma_\mu,\gamma_\nu\}=2 \eta_{\mu \nu} I
\end{equation}
where $I$ is the $2 \times 2$ identity matrix and $\eta_{\mu \nu}=\textrm{diag}(1,-1,-1)$ is the Minkowski metric in $2+1$ space-time dimensions. 

In the Lagrangian of Eq.\eqref{eq:Lagrangian} we introduced the coupling to a gauge field $a_\mu$ through the covariant derivatives 
\begin{equation}
D_\mu=\partial_\mu-i a_\mu
\label{eq:covariant-derivative}
\end{equation}
The coupling to a gauge field is needed both to describe the interactions with an external electromagnetic field $A_\mu$ and also to express the charge 
currents of the fermions in terms of a dual gauge field. This latter procedure leads to a hydrodynamic theory of the Chern insulating phase.\cite{atma}

The hydrodynamic theory is derived using the procedure of functional bosonization of 
Ref. [\onlinecite{Fradkin1994}] and expanded in Ref. [\onlinecite{LeGuillou1997}] (see also Ref.[\onlinecite{Burgess1994}]). 
Following the work of Chan {\it et al.},\cite{atma} we will derive the effective hydrodynamic theory by  considering  the partition function of the fermionic theory 
with the Lagrangian of Eq.\eqref{eq:Lagrangian} coupled to a dynamical gauge field $a_\mu$ whose field strength 
$\mathcal{F}_{\mu \nu}=\partial_\mu a_\nu-\partial_\nu a_\mu$ vanishes everywhere (in space and time), 
and hence is a gauge transformation. 
For a system with periodic boundary conditions, integrating the partition function over all gauge transformations  (including large gauge transformations) 
amounts to averaging the partition function (and hence all its observables) over the torus of boundary conditions. 

The averaged partition function is
\begin{align}
Z[A_\mu]=&\nonumber\\
\int \mathcal{D} \bar \psi  \mathcal{D}\psi & \mathcal{D}a_\mu \; \prod_{x,\mu,\nu} \delta(\mathcal{F}_{\mu \nu})
\;  \exp\Big(i \int d^3x \mathcal{L}_F [\bar \psi,\psi,A_\mu+a_\mu]\Big)
\label{eq:averagedZ}
\end{align}
where $A_\mu$ is a weak external electromagnetic field (used a s source), $\mathcal{L}_F$ is the Lagrangian of Eq.\eqref{eq:Lagrangian}. 
Using the representation of the delta function
\begin{equation}
\prod_{x,\mu,\nu}  \delta(\mathcal{F}_{\mu \nu})=\int \mathcal{D}b_\mu\; \exp(i \int d^3x \; b_\mu \epsilon^{\mu \nu \lambda} \partial_\nu a_\lambda)
\end{equation}
and the invariance of the measure under shifts $a_\mu \to a_\mu-A_\mu$, 
we find that the averaged partition function can be written in the equivalent form
\begin{align}
Z[A_\mu]=&\nonumber\\
\int \mathcal{D} \bar \psi \mathcal{D}\psi & \mathcal{D}a_\mu \mathcal{D}b_\mu \;
\exp\Big(i \int d^3x \mathcal{L}[\bar \psi,\psi,A_\mu,a_\mu,b_\mu]\Big)
\label{eq:Zaveraged+b}
\end{align}
The Lagrangian in the exponent of Eq.\eqref{eq:Zaveraged+b} is given by
\begin{equation}
\mathcal{L}[\bar \psi,\psi,A_\mu,a_\mu,b_\mu]=b_\mu \epsilon^{\mu \nu \lambda}\partial_\nu \left(a_\lambda-A_\lambda\right)+
\mathcal{L}_F[\bar \psi,\psi,a_\mu]
\label{eq:Laveraged}
\end{equation}
where the Lagrangian $\mathcal{L}_F$  on the r.h.s. of Eq.\eqref{eq:Laveraged} is given in Eq.\eqref{eq:Lagrangian}. 
In the Chern insulating phase, this expression leads to the $BF$ topological field theory form of the hydrodynamic theory.\cite{atma,Cho2011}

It is now straightforward to show\cite{Fradkin1994,LeGuillou1997,atma} that the fermionic currents $j_\mu$ can be expressed in terms of the dual 
hydrodynamic field $j_\mu\equiv \epsilon_{\mu \nu \lambda} \partial^\nu b^\lambda$ 
as an operator identity.
This hydrodynamic identity is the starting point of the effective field theory of the fractional quantum Hall fluids.\cite{Frohlich1991b,Wen1992,Wen1995}

On the other hand the conserved and gauge-invariant fermionic currents $j_\mu$ have the explicit form
\begin{align}
j_0=&\frac{\delta \mathcal{L}_F}{\delta a_0}=\bar \psi_a \gamma_0 \psi_a=\psi_a^\dagger \psi_a \label{eq:j0}\\
j_1=&\frac{\delta \mathcal{L}_F}{\delta a_1}=i \bar \psi_a \Big(\gamma_1 D_1+\gamma_2 D_2\Big)\psi_a+\textrm{h.c.} \label{eq:j1}\\
j_2=&\frac{\delta \mathcal{L}_F}{\delta a_2}=i \bar \psi_a \Big(-\gamma_1 D_2+\gamma_2 D_1\Big)\psi_a+\textrm{h.c.}\label{eq:j2}
\end{align}
where $D_1$ and $D_2$ denote the spatial components of the covariant derivative, and where the summation over the index $a$ has been assumed. 
Notice that, unlike the relativistic Dirac theory but in close resemblance to the non-relativistic case, 
the spatial components of the fermionic current depend explicitly on the gauge field $a_\mu$, as expected for a theory with dynamical exponent $z=2$.

We will now proceed to derive an effective action which is accurate in the large $N$ limit (but which is qualitatively correct for all finite $N$). 
To this end we will decouple the four-fermion interactions in the Lagrangian $\mathcal{L}_F$ by means of a Hubbard-Stratonovich transformation. 
In terms of three real Hubbard-Stratonovich 
fields $M_0(x)$, which couples to the time-reversal-symmetry-breaking order parameter $\Phi_0$ (of Eq.\eqref{eq:time-reversal-order-parameter}), 
and $M_1(x)$ and $M_2(x)$, which couple to the components of the nematic order parameter ${\bm \Phi}$ (of Eq.\eqref{eq:nematic-order-parameter}),
the Lagrangian $\mathcal{L}_F$ of Eq.\eqref{eq:Lagrangian} takes the form
\begin{widetext}
\begin{equation}
\mathcal{L}_F[\bar \psi,\psi,a_\mu,M_0,\bm{M}]=
\bar \psi_a (x) \Big(i \gamma_0 D_0- \gamma_1(D_1^2-D_2^2)- \gamma_2 (D_1 D_2+D_2D_1)+M_0(x)+\bm{M}(x) \cdot \bm{\gamma} \Big) \psi_a(x)
- \frac{N}{2g_0} M_0(x)^2-\frac{N}{2g} \bm{M}^2(x)
\label{eq:Lagrangian+HS}
\end{equation}
\end{widetext}
Upon integrating-out the fermionic fields we obtain the following expression for the averaged partition function 
\begin{align}
Z[A_\mu]=&\nonumber\\
 \int  \mathcal{D}a_\mu & \mathcal{D}b_\mu  Z[a_\mu] 
 \exp\Big(i \int d^3x \; N\; b_\mu \epsilon^{\mu \nu \lambda}\partial_\nu \left(a_\lambda-A_\lambda\right)\Big)
\end{align}
where we  scaled the $b_\mu$ field by a factor of $N$ for future convenience. The partition function $Z[a_\mu]$ is given by
\begin{equation} 
 Z[a_\mu]=
 \int \mathcal{D}M_0 \mathcal{D}\bm{M} \exp(i N S[a_\mu,M_0,\bm{M}])
 \label{eq:ZcalA}
\end{equation}
where 
\begin{align}
S[a_\mu, M_0,\bm{M}]&=\nonumber\\
-\int d^3x & \left[ \frac{1}{2g_0} M_0^2(x)+\frac{1}{2g} \bm{M}(x)^2\right]
\nonumber\\
&-i \textrm{Tr} \ln   \mathcal{M}[a_\mu,M_0,\bm{M}]
\label{eq:effective-action}
\end{align}
and $\mathcal{M}$ is the differential operator
\begin{widetext}
\begin{equation}
\mathcal{M}[a_\mu,M_0,\bm{M}]=i \gamma_0 D_0-  \gamma_1(D_1^2-D_2^2)- \gamma_2 (D_1 D_2+D_2D_1)+M_0(x)+\bm{M}(x) \cdot \; \bm{\gamma} 
\label{diffop}
\end{equation}
\end{widetext}
is the  action used in Eq.\eqref{eq:ZcalA}. Notice that the Hubbard-Stratonovich fields $M_0$ and $\bm{M}$ have units of 
$(\textrm{momentum})^2\equiv \textrm{energy}$ 
(which is consistent since $z=2$.)

 Putting it all together we find that the partition function of the full problem is
\begin{equation}
 Z[A_\mu]=\int \mathcal{D} b_\mu \mathcal{D} a_\mu \mathcal{D} M_0 \mathcal{D} \bm{M} e^{i N S_{\rm eff}[a_\mu,M_0,{\bm M},A_\mu]}
 \end{equation}
 where the effective action is
 \begin{equation}
 S_{\rm eff}=S[a_\mu, M_0,\bm{M}]+\int d^3x \;  b_\mu \epsilon^{\mu \nu \lambda}\partial_\nu \left(a_\lambda-A_\lambda\right)
\label{effective-action-total}
\end{equation}
Here $S[a_\mu, M_0,\bm{M}]$ is given by Eq.\eqref{eq:effective-action}.
Notice that from Eq.\eqref{eq:Lagrangian+HS} the following identities hold
\begin{equation}
\frac{\Phi_0}{N} = \frac{M_0}{g_0}, \qquad
\frac{{\bm \Phi}}{N} =\frac{\bm{M}}{g}
\label{eq:constraints}
\end{equation}
As usual, the correlation functions of the Hubbard-Stratonovich fields are (essentially) the same as those of the order parameters. 

We can now proceed to solve this theory in the large $N$ limit. 
The effective action we are seeking will be obtained in the leading order of the $1/N$ expansion which is equivalent to a one-loop approximation. 
(For a general discussion of large $N$ (``vector'') field theories see, e.g., the extensive review of Ref.[\onlinecite{Moshe2003}].) 

In the large $N$ limit the partition function $Z[a_\mu]$ (of Eq.\eqref{eq:ZcalA}) is well approximated by an expansion 
about the saddle-points of the effective action $S_{\rm eff}$ of Eq.\eqref{eq:effective-action}. 
Here we will seek translationally-invariant states, such as the phases with spontaneously broken time-reversal invariance, with $\langle \Phi_0\rangle \neq 0$, 
and/or spontaneously broken rotational invariance, with $\langle  {\bm \Phi} \rangle \neq 0$. 
In what follows the gauge field $a_\mu$ can be taken to be a weak perturbation (and hence it will not affect the saddle-point equations). Hence we will set $a_\mu=0$ in the saddle-point equations. The effects of quantum fluctuations of the gauge field $a_\mu$ will appear in the $1/N$ corrections.

The saddle-point-equations (the ``gap equations'') are
\begin{align}
 \frac{\delta S_{\rm eff}}{\delta M_0(x)}=0 & \Rightarrow & \frac{m}{g_0}=& -i \; \textrm{tr} G(x,x;m,\bm{M})\\
  \frac{\delta S_{\rm eff}}{\delta \bm{M}(x)}=0 & \Rightarrow &\frac{\bm{M}}{g}=& -i  \;\textrm{tr} \left[G (x,x;m,\bm{M}) \bm{\gamma}\right]
\label{eq:gap-equations}
\end{align}
\label{eq:SPE}
where a sum over repeated indices is assumed and the trace runs over the spinor indices.
$S_{\alpha \beta}(x,x';m, \bm{M})$ (with $\alpha,\beta=1,2$ being the spinor indices) is the Feynman (time-ordered) 
propagator of a fermionic field with $z=2$ 
with constant values of the fields $M_0\equiv m$  and $\bm{M}$,
\begin{widetext}
\begin{equation}
G_{\alpha \beta}(x,x'; m, \bm{M})=-i \langle T(\psi_\alpha(x) \bar \psi_\beta(x'))\rangle
= \langle x, \alpha \vert \Big(i \gamma_0 \partial_0- \gamma_1(\partial_1^2-\partial_2^2)-  \gamma_2 2 \partial_1 \partial_2+m+\bm{M} \cdot \; \bm{\gamma} \Big)^{-1}\vert x', \beta \rangle
\label{eq:Feynman-propagator}
\end{equation}
In frequency and momentum space $q_\mu=(q_0,\bm{q})$, the Feynman propagator is (dropping the spinor indices)
\begin{equation}
G(p;m, {\bm M})=\frac{1}{p_0\gamma_0-(p_1^2-p_2^2)\gamma_1-2p_1p_2 \gamma_2-m -\bm{M} \cdot \bm{\gamma}-i\epsilon}
\label{eq:Feynman-propagator-pmu}
\end{equation}
\end{widetext}
from where we read-off the spectrum of (one-particle) fermionic excitations 
\begin{equation}
E_\pm (\bm{q};m,\bm{M})=\pm E(\bm{q};m,\bm{M})
\label{eq:one-particle-spectrum}
\end{equation}
and 
\begin{equation}
E(\bm{q};m,\bm{M})= \sqrt{\left(q_1^2-q_2^2+M_1\right)^2+\left(2q_1q_2+M_2\right)^2+m^2}
\label{eq:E(q)}
\end{equation}
Clearly, $M_0=m$ is  a (time-reversal symmetry breaking) mass gap, 
and $\bm{M}$ breaks rotational invariance, by splitting the QBC into two Dirac cones, along a direction and by an amount set by $\bm{M}$.

Upon computing the traces over the spinor indices, and after an integration over frequencies, the ``gap'' equations Eq.\eqref{eq:gap-equations} can be put in the form 
\begin{align}
\frac{m}{g_0}&=
 \int \frac{d^2q}{(2\pi)^2}\;  \frac{m}{E(\bm{q};m,\bm{M})}
\label{eq:SPEm}
\\
\frac{M_1}{g}&=
 \int \frac{d^2q}{(2\pi)^2}\;  \frac{q_1^2-q_2^2+M_1}{E(\bm{q};m,\bm{M})}
\label{eq:SPEM1}
\\
\frac{M_2}{g}&=
 \int \frac{d^2q}{(2\pi)^2}\;  \frac{2q_1q_2+M_2}{E(\bm{q};m,\bm{M})}
\label{eq:SPEM2}
\end{align}
where $E(\bm{q};m,\bm{M})$ is given in Eq.\eqref{eq:E(q)}.
The integrals in Eqs. \eqref{eq:SPEm}, Eq.\eqref{eq:SPEM1} and Eq.\eqref{eq:SPEM2} 
are logarithmically divergent at large momenta $\bm{q}$ 
and require a UV momentum cutoff $\Lambda \sim \pi/a$, where $a$ is the lattice spacing. 
This logarithmic divergence is a consequence of the marginally relevant nature of the interactions.

In the $N \to \infty$ limit, the ground state energy density of the system $\mathcal{E}(m,\bm{M})$ is 
\begin{equation}
 \mathcal{E}(m,\bm{M})=\frac{N}{2g_0} m^2+\frac{N }{2g} \bm{M}^2-N \int \frac{d^2q}{(2\pi)^2} E(\bm{q};m,\bm{M})
\label{eq:SPE-gnd}
\end{equation}
where we have filled up the negative energy states. This ground state energy density has extrema at the values of $m$ and $\bm{M}$ which are the simultaneous 
 solutions of Eq.\eqref{eq:SPEm}, Eq.\eqref{eq:SPEM1}, and Eq.\eqref{eq:SPEM2}.

The saddle-point equations, Eq.\eqref{eq:SPEm}, Eq.\eqref{eq:SPEM1}, and Eq.\eqref{eq:SPEM2}, have three types of uniform solutions: 
a) an isotropic (or $C_4$ invariant) phase with $m\neq0$ and $\bm{M}=0$ in which time reversal invariance is spontaneously broken which is 
an insulating (Mott)  phase with a spontaneous QAH effect, 
b) a phase with $m=0$ but with $\bm{M} \neq 0$ with a spontaneously broken rotational (or $C_4$) invariance which is a nematic semi-metal 
with a spectrum of two massless Dirac fermions, and 
c) a coexistence phase with $m\neq 0$ and $\bm{M}\neq 0$, in which both time-reversal and rotational invariance are spontaneously broken, 
i.e. this is an insulating  nematic QAH phase. 

In Ref. [\onlinecite{kai}] it was found that, for certain range of parameters  the quantum phase transition from the QAH phase to the nematic 
QAH phase is continuous while the subsequent transition to a the nematic semimetal is first order. 
The details of the phase diagram depend also on the parameters $t$ and $t'$ , defined in the free fermion Hamiltonian of Eq.\eqref{eq:h0-QBC}, 
that break the continuous symmetry under rotations dow to the $C_4$ point-group symmetry (for the case of the checkerboard lattice).   

In this paper we will focus on the (isotropic or $C_4$-symmetric) QAH phase and its continuous 
quantum phase transition to the nematic QAH phase in which both orders are present. In the $N \to \infty$ limit the ground state energy density of the QAH phase is
\begin{equation}
\mathcal{E}(m,\bm{M})=\mathcal{E}_0+\frac{m^2}{2g_0}-\frac{m^2}{8\pi} \ln \left(\dfrac{2\Lambda^2}{|m|}\right)
\label{eq:Egnd-QAH}
\end{equation}
where $\mathcal{E}_0=-\Lambda^2/(8\pi)$ (here and below $\Lambda$ is a momentum cutoff, $\Lambda \sim \pi/a$) is the ground state energy density of free 
fermions with a QBC, and
where we have kept the leading (divergent) terms in $\Lambda^2/|m| \to \infty$. in Eq.\eqref{eq:Egnd-QAH} we have omitted an overall factor of $N$.

The ground state energy of Eq.\eqref{eq:Egnd-QAH}
is minimized if the saddle-point equation Eq.\eqref{eq:SPEm} is satisfied, which now becomes
\begin{equation}
\dfrac{1}{g_0}=\frac{1}{4\pi} \ln\left(\dfrac{2\Lambda^2}{|m|}\right)
\label{eq:SPEm'}
\end{equation}
The solution of this equation is
\begin{equation}
|m|=2\Lambda^2 \; \exp\left(-\frac{4\pi}{g_0}\right)
\label{eq:gap}
\end{equation}
which has the characteristic form of a marginally relevant perturbation.
From now on we will assume that the leading instability of the system is to the QAH phase, 
which opens the finite gap $m$ is the fermion spectrum and breaks spontaneously time-reversal invariance. 

We will consider the case in which the onset of nematic order takes place inside the QAH phase. In this situation the nematic order will be weak and its onset will not affect appreciably, to lowest order, the time-reversal-symmetry breaking mass gap $m$. 
With these assumptions we can expand the ground state energy of Eq.\eqref{eq:SPE-gnd} in powers of the nematic order parameter $\bm{M}$ up to quartic order, 
which has the form
\begin{equation}
\mathcal{E}(m,\bm{M})=\mathcal{E}_{QAH}+r(m) \bm{M}^2+ u(m) \bm{M}^4+O(\bm{M}^6)
\label{eq:E_g-m+M}
\end{equation}
where $\mathcal{E}_{QAH}$
 is the ground state energy of the nematic phase, and  the parameters $r(m)$ and $u(m)$ are
\begin{equation}
r(m)=\frac{1}{2g}-\frac{1}{8\pi} \ln \left(\dfrac{2\Lambda^2}{|m|}\right), \qquad u=\dfrac{21}{256 \pi} \frac{1}{m^2}
\label{eq:gnd-expansion}
\end{equation}  
From here we find that there is a (quantum) phase transition to a nematic QAH phase at a critical value $g_c$,
\begin{equation}
\frac{1}{g_c}=\frac{1}{4\pi} \ln \left(\frac{2\Lambda^2}{|m|}\right)
\end{equation}
Within these approximations, the transition takes place at $g_c = g_0$. For $g>g_c$ nematic order parameter $\bm{M}$ has a non-vanishing expectation value, 
\begin{equation}
\bar{|\bm{M}|}=\left(\frac{-r(m)}{2u(m)}\right)^{1/2}=  A\; |m|\; \left(\frac{1}{g_c}-\frac{1}{g}\right)^{1/2} 
\label{eq:|M|} 
\end{equation}             
where $A^2=64 \pi/21$. Further inside the nematic QAH phase the QAH order parameter, $m$, 
becomes progressively suppressed until a first-order quantum phase transition to a nematic semimetal phase is reached.\cite{kai}

\section{Effective action and $1/N$ expansion}
\label{sec:1/N}

We will now derive the effective field theory for the quantum fluctuations in the QAH phase close to the nematic quantum phase transition. 
To this end we will compute the effects of quantum fluctuations to the lowest order in the $1/N$ expansion. 
In the QAH phase the only field with a non-vanishing expectation value is the field $M_0$, whereas the nematic field $\bm{M}$ has a vanishing expectation 
value in the QAH phase (but not in the nematic phase). 
By gauge invariance the gauge fields $a_\mu$ and $b_\mu$ cannot have a non-vanishing expectation value (although their fluxes could).

The fluctuations of the time-reversal symmetry-breaking field $M_0$ are massive in the QAH phase (and in the nematic QAH phase). 
Since we are interested in the effective field theory close to the transition to the nematic QAH phase we will not be interested in the fluctuations of this massive field, 
whose main effect is a renormalization of the effective parameters. 
Thus in what follows we will ignore the fluctuations of the field $M_0$ about the $N=\infty$ expectation value $M_0=m$.

We will now expand the effective action of Eq.\eqref{eq:effective-action} to lowest orders in the $1/N$ expansion. Let us denote by $G_0(x,x';m)$ 
\begin{equation}
G_0(x,x'; m)\equiv \langle x  \left|  \mathcal{M}^{-1}_0 \right|x' \rangle
\label{eq:calM0}
\end{equation}
the Feynman propagator of the fermions in the QAH phase given by Eq.\eqref{eq:Feynman-propagator}.
Here we implicit the spinor indices and set the expectation value of the nematic field $\bm{M}$ to zero and $M_0=m$. In Eq.\eqref{eq:calM0} 
 $\mathcal{M}_0$ is the differential operator of Eq.\eqref{diffop} in the symmetric phase with broken time reversal symmetry. 
 
 In momentum space the propagator of Eq.\eqref{eq:propagator} becomes
\begin{equation}
G_0(p)=\frac{p_0 \gamma_0-(p_1^2-p_2^2)\gamma_1-2p_1p_2\gamma_2+m}{p_0^2-(p_1^2+p_2^2)^2-m^2-i\epsilon}
\label{eq:propagator}
\end{equation}

 The expansion in powers of $1/N$ can now be determined by using the expansion of the logarithm 

\begin{align}
\textrm{tr} \ln \mathcal{M}=&\textrm{tr} \ln \left(\mathcal{M}_0+\delta \mathcal{M}\right)\nonumber\\
=&\textrm{tr} \ln \mathcal{M}_0+\textrm{tr} \ln \left(I+\mathcal{M}_0^{-1}\delta \mathcal{M}\right)
\end{align}
where
\begin{align}
\textrm{tr}\;  \ln &\left(I+\mathcal{M}_0^{-1}\delta \mathcal{M}\right)= \nonumber\\
\textrm{tr}\; & \left(\mathcal{M}_0^{-1} \delta \mathcal{M}\right) -\frac{1}{2} \textrm{tr}\;  \left(\mathcal{M}_0^{-1} \delta \mathcal{M}\right)^2+&\frac{1}{3}  \textrm{tr}\;  \left(\mathcal{M}_0^{-1} \delta \mathcal{M}\right)^3\nonumber\\
&+\ldots
\label{eq:expansion-determinant}
\end{align}
where $\mathcal{M}_0$ and $\delta \mathcal{M}$ are the operators
\begin{align}
\mathcal{M}_0=&i \gamma_0 \partial_0-  \gamma_1(\partial_1^2-\partial_2^2)- \gamma_2 2 \partial_1 \partial_2 +m \nonumber\\
\delta \mathcal{M}=&\bm{M}(x) \cdot \; \bm{\gamma} +a_{\mu} \mathcal{J}^{\mu}-{\bm T}_{ij} a_i a_j
\end{align}
with the vertices $\mathcal{J}_{\mu}$ and  given by  
\begin{align}
\mathcal{J}_0=&\gamma_0  \\
\mathcal{J}_1=& i \gamma_1 \partial_1+i \gamma_2 \partial_2+\textrm{h.c.} \\
\mathcal{J}_2=&-i \gamma_1 \partial_2+i \gamma_2 \partial_1+\textrm{h.c.}  
\label{eq:vertices1}
\end{align}
where $i,j=1,2$ label the two spatial components of the gauge field $a_\mu$, and the matrix ${\bm T}$ is
\begin{equation}
{\bm T}=
\begin{pmatrix}
\gamma_1 & \gamma_2\\
\gamma_2 & -\gamma_1
\end{pmatrix}
\label{eq: vertices2}
\end{equation}
where  $\gamma_1$ and $\gamma_2$ are the two spatial Dirac gamma matrices.

The terms in the expansion of Eq.\eqref{eq:expansion-determinant}
 that are quadratic in the nematic fields $\bm M$  and on the hydrodynamic gauge field $a_\mu$ represent the leading quantum fluctuations about the $N=\infty$ limit.  The effective action for the quantum fluctuations of the hydrodynamic gauge field $a_\mu$ and the nematic fields $\bm{M}$ have the form
 
\begin{equation}
S_{\rm eff}[a_\mu,\bm{M}]=S_{\rm eff} [a_\mu]+S_{\rm eff}[\bm{M}]+S_{\rm eff}[a_\mu,\bm{M}]
\end{equation}

Here $S[\bm{M}]$ describes the dynamics of the nematic field, and will be studied in detail in the next section. In this section, we focus on the effective action of the hydrodynamic gauge fields and on their coupling to the nematic fields, $S_{\rm eff}[a_\mu]+S_{\rm eff}[a_\mu,\bm{M}]$. The details of the Feynman diagrams and of the calculations included in this section can be found in Appendix \ref{app:effective-gauge-theory}. 
The resulting effective Lagrangian is
\begin{widetext}
\begin{align}
\mathcal{L}_{\rm eff}[a_\mu]+\mathcal{L}_{\rm eff}[a_\mu,\bm{M}]=&
 \frac{N}{4\pi}\epsilon^{\mu \nu \rho} a_{\mu} \partial_{\nu} a_{\rho}+N b_{\mu}\epsilon^{\mu \nu \rho}  \partial_{\nu} (a_{\rho}-A_{\rho}) \nonumber\\
&+\frac{N}{8\pi}  \left(\frac{1}{m} + \frac{M_1}{2m^2}\right)(\partial_0 a_1-\partial_1 a_0)^2  +\frac{N}{8\pi} \left(\frac{1}{m}-\frac{M_1}{2m^2}\right)(\partial_0 a_2-\partial_2 a_0)^2 \nonumber \\
 &+ \frac{N}{8\pi} \frac{M_2}{m^2} (\partial_0 a_1-\partial_1 a_0)(\partial_0 a_2-\partial_2 a_0)-\frac{N}{4\pi}(\partial_2 a_1-\partial_1 a_2)^2
  \label{eq:effective-action-gauge-fields}
 \end{align}
\end{widetext}
The effective gauge theory is a Maxwell-Chern-Simons theory. The first term is the Chern-Simons term from the nontrivial fermion band, the second term is the BF term obtained from the functional bosonization technique we used. It is straightforward to see that this effective action predicts that the QAH phase has a Hall quantized Hall conductivity $\sigma_{xy}=N e^2/h$, as expected for the quadratic band crossing case.\cite{kai}

The rest of the terms in the effective action of Eq.\eqref{eq:effective-action-gauge-fields} are the parity-even Maxwell terms and the local coupling of  the fluctuation of the nematic fields  to the hydrodynamic gauge field. The latter has the  form of an  effective spatial anisotropy. Hence, it is apparent from Eq.\eqref{eq:effective-action-gauge-fields} that the nematic order parameters  couple to the gauge fields as an effective spatial metric. To make this more clear, let us rewrite the Maxwell terms, $\mathcal{L}_{\rm Maxwell}$ in  the form (for comparison, see Ref.[\onlinecite{Barci-2011}])
 \begin{align}
\mathcal{L}_{\rm Maxwell}=&-\frac{N}{8\pi \sqrt{2m}} f_{\mu \nu} g^{\mu \alpha} f_{\alpha \beta} g^{\beta \nu}\\
f_{\mu \nu}=& \partial_\mu a_\nu-\partial_\nu a_\mu\\
 g_{\mu \nu}=&\eta_{\mu \nu}+ \frac{1}{2m} Q_{\mu \nu}
\end{align} 
where we have rescaled the time coordinate and temporal component of the gauge field $x_0 \rightarrow \frac{1}{\sqrt{2m}}x_0,a_0\rightarrow\sqrt{2m} a'_0$ so as to renormalize the dielectric constant and make the ``speed of light'' be $1$. 
The modified metric in the Maxwell term are composed of a regular flat metric of $2+1$-dimensional Minkowski space-time, $\eta_{\mu \nu}={\rm diag}(1,-1,-1)$, locally modified by a traceless metric $Q_{\mu \nu}$ induced by the local spatial anisotropy.
The traceless symmetric tensor $Q_{\mu \nu}$ only has non-vanishing spatial components,
 \begin{eqnarray}
 Q_{\mu \nu}= 
\begin{pmatrix} 
0 & 0 & 0 \\
0 & M_1& M_2 \\
0 & M_2& -M_1
\end{pmatrix}
\end{eqnarray} 

From the expression of $Q_{\mu \nu}$, it is clear that this is the hydrodynamic theory of a gauge field  on a manifold with a fluctuating nontrivial (purely spatial) metric due to the coupling to the nematic field. As the fluctuation of the nematic field modifies the local metric, in the anisotropic phase, where the tensor $Q_{ij}$ (or, equivalently, $\bm{M}$) acquires an nonzero expectation  value, the Maxwell term becomes anisotropic. This leads to anisotropic transport (at finite wave vector $\bm q$) in the nematic QAH. This phenomenon is equivalent to having an anisotropic dielectric dielectric tensor that plays the role  of the metric tensor we introduced here.

\section{effective theory of the nematicity}
\label{sec:nematic-effective-action}

Let us now derive the effective theory of the nematic field ${\bm {M}}$. 
The effective action $S_{\rm eff}(\bm M)$, obtained for the integration of the fermions and from the Hubbard-Stratonovich fields, has the form
\begin{align}
S_{\rm eff}(\bm M)=N \ln \textrm{det}(G^{-1}_{0}- {\bm M} \cdot  {\bm  \gamma})  
-\int d^3x \; \frac {N}{2g} {\bm M}^2
\end{align}
By expanding the effective action to the quadratic order, we get
\begin{align}
S_{\rm eff}=&-\frac {N}{2} \textrm{tr}(G_{0} {\bm \gamma} \cdot {\bm M} G_{0} {\bm \gamma} \cdot  {\bm M}) -\int d^3x \; \frac {N}{2g} {\bm M}^2\nonumber  \\
= -\frac {N}{2} &\int \frac{d^3 p}{(2\pi)^3} M_{i}(-p)\Gamma_{ij}(p) M_{j}(p) -\frac{N}{2g} \int \frac{d^3p}{(2\pi)^3} |{\bm M}(p)|^2
\end{align}
where $\Gamma_{ij}(p)$ is the one-loop kernel 
\begin{equation}
\Gamma_{ij}(p) =\int \frac{d^3 k}{(2\pi)^3} \textrm{tr}(\gamma_{i} G_{0}(p+k)\gamma_{j} G_{0}(k))
\end{equation}
which is given by the  self-energy diagram discussed in Appendix \ref{app:effective-action-nematic}.

Let us now define a $2 \times 2$ traceless symmetric tensor field ${\bm Q}$ which is natural to describe a nematic phase  \cite{Oganesyan-2001,chaikin-1995}
\begin{eqnarray}
{\bm Q}=
\begin{pmatrix}
  M_1 & M_2 \\
  M_2 & -M_1
\end{pmatrix}
\end{eqnarray}
At long wavelengths  and low frequencies, the effective Lagrangian of the nematic order parameter $\mathcal{L}[\bm Q]$ is
\begin{align}
\frac{1}{N} \mathcal{L}_{\rm eff}[\bm Q]=& -\chi(m) \epsilon^{bc} Q_{ab} \partial_0 Q_{ac} - r(m) \textrm{Tr}[{\bm Q} {\bm Q}]\nonumber\\
+\kappa_1 \textrm{Tr}[{\bm Q}&{\bm K}{\bm Q}]+ \kappa_2\textrm{Tr}[\sigma_1 {\bm Q}{\bm K}'{\bm Q}]
-u(m) \textrm{Tr}[{\bm Q}{\bm Q}{\bm Q}{\bm Q}]
\label{eq:Leff-Q}
\end{align}
where $\bm K$ and ${\bm K}'$ are the $2 \times 2$ symmetric matrix differential operators
\begin{equation}
{\bm K}=  
\begin{pmatrix}
  \partial^2_1 &   \partial_1 \partial_2 \\
   \partial_2 \partial_1  &     \partial^2_2 
\end{pmatrix},\quad
{\bm K}'= 
\begin{pmatrix}
  \partial_2 \partial_1  &   \partial^2_2\\
   \partial^2_1 &     \partial_2 \partial_1 
\end{pmatrix}
\end{equation}
and $\sigma_1$ is the (symmetric and real) Pauli matrix.

The coefficients $r(m)$ and $u(m)$ in Eq.\eqref{eq:Leff-Q} were given already in Eq.\eqref{eq:gnd-expansion}. The coefficient $\chi(m)$ shown in Eq.\eqref{eq:Leff-Q}, is given by
\begin{equation}
\chi(m)=\frac{1}{64\pi} \frac{1}{m}
\label{eq:chi}
\end{equation}
The coefficient coefficient $\chi(m)$ depends on {\em both}  the {\em magnitude} and  the {\em sign} of the parameter $m$, i.e. on the expectation value of the order parameter that measures the spontaneous breaking of time-reversal invariance in the Mott Chern insulator. This behavior  is reminiscent of the Parity Anomaly of a Dirac fermion in $2+1$ dimensions.\cite{Deser-1982,Redlich-1984}  In the next section we will see shortly that $\chi(m)$ is related to the Hall viscosity and hall torque viscosity of the spontaneous QAH phase. Moreover, the presence of this Berry phase term makes the dynamic critical exponent of the effective theory of the nematic fields to be $z=2$.

The first term of the effective action $\mathcal{L}_{\rm eff}[\bm Q]$ of Eq.\eqref{eq:Leff-Q} is of first order in time derivatives, reflecting the spontaneous breaking of time-reversal invariance in the (spontaneous) QAH phase and, hence,
  is {\em odd} under time-reversal. This term can be regarded as a Berry phase of the time evolution of the nematic order parameter field. Maciejko and collaborators\cite{Maciejko-2013} have shown that it is possible to rewrite the effective field theory of the nematic order parameter field as a non-linear sigma model whose target space is a hyperbolic space, a coset of $SO(2,1)$. The form of our Berry phase term is consistent with the one discussed by of Maciejko and collaborators\cite{Maciejko-2013}  in the limit ${\bm Q}\ll 1$ which we have used here.
  
  Before we discuss the phases of this theory and the behavior of the nematic degrees of freedom it is worth to comment on the symmetries of the effective Lagrangian of Eq.\eqref{eq:Leff-Q}. As it is apparent this effective Lagrangian is invariant under a global rotation of the nematic order parameter field (modulo $\pi$). This symmetry is the result of setting $t=t'$ in the lattice Hamiltonian of Eq.\eqref{eq:h0-QBC} and of the fact that we kept only the lowest terms in momenta in the long wavelength theory of the fermions of Eq.\eqref{eq:H0-long-wavelengths}. On the other hand, if $t\neq t'$ the  effective low-energy theory has a lower $C_4$ symmetry. At the level of the nematic order parameter, this is equivalent to an Ising symmetry (of rotations by $\pi/2$. The same type of symmetry breaking is obtained in the corrections to Eq.\eqref{eq:H0-long-wavelengths} of order $p^4$ (or higher) in the effective low-energy Hamiltonian of the fermions. The net effect of these corrections are nominally irrelevant operators which break the continuous $O(2)$ symmetry down to a discrete (Ising) symmetry.  

\subsection{The isotropic QAH phase}

In the isotropic QAH phase, and to lowest order in the $1/N$ expansion, we find that the stiffnesses are 
\begin{equation}
\kappa_1= \frac{1}{12 \pi |m|}, \qquad \kappa_2=0
\label{eq:stiffnesses}
\end{equation}
Hence, in the isotropic phase, the terms of the effective action that depend on the spatial gradients, after an integration by parts,  can be written in the form
\begin{equation}
-\kappa_1\textrm{Tr}[{\bm Q}{\bm K}{\bm Q}]= \kappa_1((\nabla\cdot\bm{M})^2+(\nabla\times\bm{M})^2)
\end{equation}
Hence the two Frank constants are equal in the isotropic phase. 

It is straightforward to see that the nematic modes are gapped in the isotropic phase and that their gap vanishes at the quantum phase transition. Again, provided the explicit lattice symmetry breaking effects we discussed above can be neglected, the spectrum of nematic modes will ge gapped but degenerate.

\subsection{The nematic QAH phase}

However  in the nematic QAH phase where, the rotational symmetry is spontaneously broken. This has two consequences. One is that instead of a single Frank constant (stiffness) we now find two,
\begin{equation}
\kappa_1= \frac{1}{12 \pi |m|},\qquad \kappa_2=\frac{|\bar {\bm M}|}{16 \pi m^2}
\label{eq:Frank-nematic}
\end{equation}
where $\bm Q$ represents now the {\em fluctuations} of the nematic order parameter in the nematic QAH phase, 
$|\bar {\bm M}|$ is the expectation value of the nematic field in the $N \to \infty$ limit and is  given in Eq.\eqref{eq:|M|}.  By symmetry, the Frank stiffness $\kappa_2$ is an odd function of the magnitude of the nematic order parameter $|\bar {\bf M}|$. Thus,  provided we restrict ourselves to the vicinity of the transition, in Eq.\eqref{eq:Frank-nematic}  we may keep only the leading (linear) term.

 Hence, as expected, in the nematic QAH phase there are two Frank constants, and
the spatial terms of the effective Lagrangian for the nematic fluctuations now becomes (also after an integration by parts)
\begin{align}
&-\kappa_1\textrm{Tr}[{\bm Q}{\bm K}{\bm Q}]- \kappa_2\textrm{Tr}[\sigma_1 {\bm Q}{\bm K}'{\bm Q}]\nonumber\\
&= (\kappa_1+\kappa_2)(\nabla\cdot{\bm M})^2+(\kappa_1-\kappa_2)(\nabla\times{\bm M})^2
\label{eq:nematic-energy}
\end{align}
which is the generally expected form for the energy of nematic fluctuations.\cite{deGennes-1993,chaikin-1995} 
A similar result generally holds in other electronic nematic phases.\cite{Oganesyan-2001}

The other consequence is that there is a gapless Goldstone mode of the spontaneously broken symmetry. Again, if the microscopic theory only has a discrete $C_4$ invariance  the Goldstone modes is gapped but the gap can be small if the explicit symmetry breaking is weak.

\subsection{Critical Behavior}

We will now discuss briefly the critical behavior. By examining the effective Lagrangian of Eq.\eqref{eq:Leff-Q} 
we see that the nematic order parameter field has scaling dimension $1$, i.e. $[Q]=l^{-1}$ (where $l$ is a length scale) or 
$\Delta_Q=1$. This scaling follows from the presence of the Berry phase term in the effective Lagrangian. Incidentally, the main effect of the Berry phase term is to make the two components of the nematic order parameter field to be canonically conjugate pairs. From the fact that the order parameter has scaling dimension $\Delta_1=1$ it follows that the scaling dimension of the quartic term of the effective Lagrangian has dimension $\Delta_4=4$ and that the effective coupling constant can be made dimensionless (by absorbing the Berry phase $\chi(m)$ in a rescaling of the nematic field). This is consistent with the fact that the dynamical exponent is $z=2$ and the the dimensionality of space is $d=2$. Hence the ``effective dimension is $d+z=4$. hence the quartic term of the Lagrangian is superficially marginal at the nematic quantum critical free field point, $r=0$. Thus, this theory appears to behave much in the same way as conventional (relativistically invariant) $\phi^4$ quantum field theory of four space-time dimensions. 

Just as in conventional $\phi^4$ theory,  the quartic term is also marginally irrelevant at the free field fixed point with $z=2$. Provided this assumption (which we have not verified) is correct, we deduce that the quantum critical behavior is that of the effective classical theory, of Eq.\eqref{eq:Leff-Q}, with logarithmic corrections to scaling. On the other hand, if the quartic term were to be marginally relevant, it would turn this quantum phase transition in to a fluctuation-induced first order transition.

Finally this theory has  a finite-temperature thermodynamic phase transition at a $T_c$ at which the nematic order is lost. If the symmetry is $O(2)$ then we expect a conventional nematic continuous (Kosterlitz-Thouless) phase transition. On the other hand if the symmetry is broken (microscopically) down to a discrete Ising ($\mathbb{Z}_2$) symmetry, the the finite-temperature transition would be in the 2D ising universality class.

\section{Transverse dissipationless response to shear stress: Hall torque viscosity in the quantum anomalous Hall state}
\label{sec:torque-viscosity}

Quantum Hall fluids and  other two-dimensional systems with broken time-reversal invariance such as Chern insulators, show a variety of dissipationless responses to external fields which do not exist in normal fluids. In a system with broken time-reversal invariance due either to an external perpendicular magnetic field or to topologically non-trivial band structures, an in-plane electric field induces a Hall current which is perpendicular to the applied  field and has  a Hall conductance which is precisely determined by the topological properties of these fluids. 
Similarly, in a two-dimensional system with broken time reversal invariance and parity, by shearing the system in one direction a momentum transfer is induced in the perpendicular direction.
As a result, the  stress tensor has an anti-symmetric component which is proportional to the shear rate. The associated transport coefficient is the  Hall viscosity.\cite{avron1995,Read-2009,taylor,vis,Hughes-2013}  

While the resulting Hall conductance is dimensionless and universal (in units of $e^2/h$), the Hall viscosity has units of length$^{-2}$. If the system is  Galilean invariant (which is the case, to a good approximation, in the 2DEG in AlAs-GaAs heterostructures and quantum wells) then the length scale is supplied by the magnetic length and, in this sense, the Hall viscosity is also universal.\cite{vis} On the other hand, in the case of topological Chern insulators, although there is a finite Hall viscosity in general it is the sums of a non-universal term (which is determined by microscopic physics) and an essentially universal term.\cite{taylor}

In this section we will first derive an expression of the Hall viscosity for the system at hand, a Chern insulator originating from an instability of a system with a  quadratic band crossing. Here we will show that the Hall viscosity is related to both the Hall conductivity of the QAH phase and with the coefficient $\chi$ of he Berry phase term obtained in Eq.\eqref{eq:effective-action-gauge-fields}. We will also see how this is related to the concept of Hall torque viscosity which we introduce below.

For a parity violating system, such as the quantum Hall fluids of 2DEGs, a change in  the background metric $g_{ij}$ of the surface on which the electron fluid resides modifies the definition of the momentum of the electrons through their coupling to the metric. A consequence of the breaking of time reversal and parity (either explicit or spontaneous) the effective field theory of the weak perturbation of the metric contains a term which is odd under parity and time reversal. Such Chern-Simons-type terms are first order in time derivatives, and their coefficient is the Hall viscosity.

  On the other hand, the fermion field of the system we are interested in is a theory of  two-component spinors and it is not Galilean invariant. A system of spinors, such as the one given in the effective long wavelength Hamiltonian of Eq.\eqref{eq:H0-long-wavelengths}, is defined with respect to a frame of orthonormal  two-component vectors ${\bm e}^a$ (with $a=1,2$) tangent to the two-dimensional space. Microscopically these vectors are tied to the local geometry of the underlying two-dimensional lattice. Thus, under a lattice deformation (which includes local rotations), these local frames, which following tradition we will call zweibeins,  accordingly change slowly. 
  
  Let us now suppose that we rotate the "spinor frame'' of the fermion field, i.e. that we make a local change of basis of the spinors. A global change of basis with a rotation axis normal to the plane is a symmetry since it is equivalent to a  rotation of the space axis. However spinor rotations about arbitrary axis and/or under a local change of basis, i.e. a change of the local frame, are not symmetries of the system. As a result of such transformations the system generally experiences a torque viscosity which is perpendicular to the axis of rotation. In what follows we will be interested in adiabatic changes in the frames of the spinors and in the Berry phase terms they induce.

We will now show that the coefficient $\chi(m)$ of the effective action of the nematic order parameter fields is related to the  Hall viscosity in the QAH phase.\cite{avron1995,Read-2009,taylor,vis} An excellent discussion of the Hall viscosity can be found in the recent work of Hughes, Leigh and Parrikar\cite{Hughes-2013} whose methods we use here.

In order to represent the local deformations of the space one couples  the frames (the zweibiens) directly to the covariant derivative. However, in our case there is an orbital degree of freedom and an analog of a spin connection is required. 
The long-wavelength Lagrangian for the free fermions on the undistorted lattice is
\begin{align}
\mathcal{L}=
\bar \psi_a (x) (i \gamma_0 \partial_0- \gamma_1(\partial_1^2-\partial_2^2)- \gamma_2 2 \partial_1 \partial_2+M_0 ) \psi_a(x)
\label{eq:QBT}
\end{align}
In this Section we will discuss the behavior of the Hall viscosity and the Hall torque viscosity in the isotropic QAH in the $N \to \infty$ limit. In this limit, and in this phase, the nematic order parameter field has vanishing expectation value and does not contribute. However its fluctuations do contribute (to order $1/N$) to the corrections at small but finite momenta of these quantities.

By adding the background distortion connecting between real space (or momentum) and orbital space, the new Lagrangian, which now depends explicitly on the frame fields ${\bm e}^a(x)$,  becomes
\begin{equation}
\mathcal{L}=
\bar \psi_{\alpha} (x) \left( i \gamma_0 \partial_0 -  T^{ij}_a e^a_k \gamma_k \partial_i \partial_j + M_0  \right)_{\alpha, \beta} \psi_{\beta}(x)
\label{eq:L-distorted}
\end{equation}
where, $a=1,2$, $\alpha,\beta=1,2$,  and $i, j, k=1,2$. As before, we have set $T_1= \sigma_z$ and $ T_2= \sigma_x$. 
The metric tensor of the 2D distorted space is $g_{ij}=e^a_i e^a_j$.
For a system on a flat metric, i.e. an undistorted lattice, the frame vectors are $e^a_i=\delta^a_i$ and, in this case, $g_{ij}=\delta_{ij} $, 
and the Lagrangian of Eq.\eqref{eq:L-distorted} reduces to our original free fermion Lagrangian of Eq.\eqref{eq:QBT}.

Here we will be interested in shear distortions and rotations, which are area-preserving diffeomorphisms. We can parametrize the frame fields ${\bm e}^a$ 
as follows
\begin{equation}
e^1_1-1=-(e^2_2-1)=e_1,\qquad  e^1_2=e^2_1=e_2
\label{eq:frame-change}
\end{equation}
Under this distortion, the free-fermion Lagrangian becomes
\begin{align}
{\cal L}=&\;
\bar \psi (x) \left( i \gamma_0 \partial_0 - \gamma_1 (\partial_1^2 - \partial_2^2) - \gamma_2 2 \partial_1 \partial_2 -M_0\right) \psi(x)
\nonumber\\
+& \; \bar \psi (x) \left(- e_1 \gamma_1 (\partial_1^2 - \partial_2^2)+e_1 \gamma_2 2 \partial_1 \partial_2 \right) \psi(x)
\nonumber\\
+& \; \bar \psi (x) \left(- e_2 \gamma_2 (\partial_1^2 - \partial_2^2) - e_2 \gamma_1 \partial_1 \partial_2 \right) \psi(x)
\label{eq:L-distorted-frames}
\end{align}
where $e_1(x)$ and $e_2(x)$ are two slowly varying functions of space and time.

After integrating-out the fermion field, the effective theory of the frame fields ${\bm e}^a$ contains a parity-violating term which appears to the first order time derivatives. In momentum and frequency space it has the form
\begin{align}
S_{\rm eff}[{\bm e}_i]=&\int \frac{d\omega}{2\pi} \frac{d^2p}{(2\pi)^2}\; i \eta({\bm p},\omega) \omega \; \epsilon^{ij} e_i ({\bm p},\omega)  e_j (-{\bm p},-\omega)\nonumber\\
& +\ldots
\end{align}
where $\eta({\bm p},\omega)$ is given by 
\begin{equation}
\eta({\bm p},\omega) =\frac{1}{i \omega} \epsilon^{ij} \frac{\delta^2 S}{ \delta e_i({\bm p},\omega)  \delta e_j(-{\bm p},-\omega) }
\end{equation}

In what follows we will only be interested in the adiabatic regime. Thus we will take the limit $\omega \to 0$.
In this limit w can expand $\eta({\bm p},0)=\eta({\bm p})$ in powers of  the momentum ${\bm p}$. In the isotropic QAH phase $\eta({\bm p})$ can only be a function of ${\bm p}^2$. To lowest orders we obtain 
\begin{equation}
\eta({\bm p})=\eta(0)+\eta_1 {\bm p}^2+ \eta_2 {\bm p}^4+\ldots
\label{eq:eta-expansion}
\end{equation}
where ${\bm p}^4=({\bm p}^2)^2$, etc.
 For symmetry reasons, only powers even powers of the momentum  are allowed to enter in this expansion. 
 
 On the other hand, in the nematic QAH insulating phase, in addition to an isotropic component of the form of Eq.\eqref{eq:eta-expansion} there is an anisotropic piece. Close to the quantum critical point the anisotropic piece of the term quadratic in momenta is a linear function of the expectation value of the nematic order parameters and has the form (up to a constant prefactor) $(p_1^2-p_2^2)M_1+2p_1p_2 M_2$. Similar considerations apply to the higher order terms in the expansion in momenta.

The zeroth-order coefficient, $\eta(0)$, in Eq.\eqref{eq:eta-expansion} is the Hall viscosity $\eta$, 
\begin{align}
\eta=&\int \frac{d^2k}{(2\pi)^2} \frac{m\;  (k_1^2+k_2^2)^2}{(k_0^2-(k_1^2+k_2^2)^2-m^2-i\epsilon)^2}  \nonumber\\
=&\frac{m}{16\pi} \ln\left(\frac{2\Lambda^2}{m}\right)-\frac{m}{16\pi} 
\label{eq:eta0}
\end{align}
which depends both on the magnitude {\em and} the sign of the mass $m$. 
Notice that the Hall viscosity, as expected, has units of $m$, or what is the same units of length$^{-2}$. The Hall viscosity $\eta=\eta(0)$ can also be computed from the correlation function of the stress tensor, $\langle T^a_iT^b_j \rangle$.\cite{taylor}

The coefficient $\eta_2$ for the term $O({\bm p}^4)$ in the expansion of Eq.\eqref{eq:eta-expansion} is proportional to the coefficient $\chi(m)$ appearing in the Berry phase term in effective nematic theory,
\begin{align}
\eta_2=&\; \epsilon^{ij} \frac{1}{i\omega}\frac{\delta^2 S}{ \delta [p^2 e_i(p)] \delta [p^2 e_j(-p)]}\nonumber\\
 \propto &\; \int \frac{d^3k}{(2\pi)^3}\frac{m}{(k_0^2-(k_1^2+k_2^2)^2-m^2-i\epsilon)^2}
 \end{align}
 Hence we find that
 \begin{align}
\eta_2 \propto &\; \lim_{\omega \to 0} \lim_{ {\bm p} \to 0} \epsilon^{bc} \frac{1}{i\omega}\frac{\delta^2 S_{\rm eff}(\bm M)}{ \delta Q_{ab}(p) \delta Q_{ac}(-p)}\nonumber\\
=&\; \chi(m)
\end{align}
Actually, the coefficient of the $p^2$ term of the expansion is proportional to the Hall conductance,
\begin{align}
\eta_1=&\; \epsilon^{ij} \frac{1}{i\omega}\frac{\delta^2 S}{ \delta [p e_i(p)] \delta [p e_j(-p)]}\nonumber\\
 \propto&\; \int \frac{d^3k}{(2\pi)^3} \frac{m \; (k_1^2+k_2^2)}{(k_0^2-(k_1^2+k_2^2)^2-m^2-i\epsilon)^2}
 \end{align}
 Hence, we also find that
 \begin{align}
\eta_1 \propto&\;\lim_{\omega \to 0} \lim_{ {\bm p} \to 0} \epsilon^{ij} \frac{1}{i\omega}\frac{\delta^2 S_{\rm eff}(\bm M)}{ \delta A_i (p) \delta A_j (-p)}\nonumber\\
=&\; \frac{1}{4}\sigma_{xy}
\end{align}

Unlike the Hall conductivity, the Hall viscosity is not a topological response as it does depend on microscopic details of the fermionic system. Furthermore, if we were to include the nematic field in Eq.\eqref{eq:QBT},  even in the isotropic phase  its fluctuations to order $1/N$ 
modify the values of  $\eta_1$ and $\eta_2$ but do not affect the value of the Hall viscosity $\eta$. 
In this sense, the relationship between $\chi $, $\sigma_{xy}$ and $\eta_1,\eta_2$ is not universal. Moreover, in the nematic phase the coefficients $\eta_1$ and $\eta_2$ become tensors, reflecting the nematic nature of the phase.

Now we come to the Hall torque viscosity. As in most (but  not all) Chern insulators, the fermion field of the quadratic band crossing model is a two component spinor which labels the two different bands. In the case of the checkerboard model the spinor labels can be traced back to the two-sublattice structure of the lattice. Suppose we now rotate the ``spinor frame'' of the fermion by  an SU(2) unitary transformation of the form
\begin{align}
\Psi'_{\alpha} (x)=\left[e^{i (-\theta_2 \sigma_x + \theta_1 \sigma_z)}\right]_{\alpha \beta}\Psi_\beta(x)
\end{align}
The rotation axis of this transformation lies on the $xz$ plane. Suppose now that we consider an infinitesimal rotation angle  so that we can expand the rotation matrix to lowest order in $\theta$,
\begin{equation}
\Psi'=\Omega \Psi, \qquad
\Omega=
\begin{pmatrix}
1-i\theta_2  &   i\theta_1\\
 i\theta_1 &    1+i\theta_2 
\end{pmatrix}
\label{eq:spinor-rotation}
\end{equation}
This is not a symmetry transformation of the Lagrangian. Indeed, 
upon  this rotation of the spinor frame, the Lagrangian Eq.\eqref{eq:QBT} changes as follows
\begin{align}
\mathcal{L}=&\;
\bar \psi' (x) (i \gamma_0 \partial_0- \gamma_1(\partial_1^2-\partial_2^2)- \gamma_2 2 \partial_1 \partial_2 -m ) \psi'(x) \nonumber\\
 - \bar \psi'& (x) (\theta_1(\partial_1^2-\partial_2^2)+\theta_2 2 \partial_1 \partial_2+m\theta_1 \gamma_1 + m\theta_2 \gamma_2) \psi'(x)
\end{align}
As we can see, the last two terms generated by a rotation of the spinor frame have exactly the same form as the nematic order parameter. in addition, the spinor  rotation also mixes with the time reversal symmetry breaking mass term (albeit with terms which are quadratic in spatial derivatives). 

It is straightforward to obtain the effective action for the spinor rotation angles in the adiabatic regime. Similarly to the calculation that we did for the Hall viscosity, here too we find  an antisymmetric term which is first order in time derivatives,
\begin{equation}
\mathcal{L}(\theta)= -\eta^s\epsilon^{ij} \theta_i \partial_0 \theta_j+\ldots
\end{equation}
where $\eta_s$ is the torque viscosity and we find it to be
\begin{equation}
\eta_s=-\frac{m}{16\pi} \ln\left(\frac{2\Lambda^2}{m}\right)+\frac{m}{8\pi} 
\label{eq:torque-viscosity}
\end{equation}
This result shows the existence of a dissipationless transport property, namely the Hall torque viscosity, which is the response of the action under an adiabatic rotation of the spinor frame.

By analogy with the stress-energy tensor for a metric distortion, here we can define the torque $\langle {\bm S} \rangle$ for the rotation of the spinor frame,
\begin{equation}
\langle S_i \rangle=\frac{\delta S}{\delta \theta_i}=A^{ij} \partial_0 \theta_j +B^{ij}  \theta_j + \ldots
\end{equation}
The second term yields the linear response between the torque and the time derivative of the rotation angle (the angular velocity). The rank tensor $A^{ij}$ is the torque viscosity. In a time-reversal and parity invariant fluid, this viscosity tensor is symmetric, indicating the rotation entails an energy cost and, furthermore, in general it is a dissipative response.
 However, in a system of spinors with broken parity and  time-reversal invariance, such as QAH phase of our system, the tensor $A^{ij}$ must have an antisymmetric part which is odd under parity. Thus, when we rotate the spinor frame in the QAH phase, there is a torque viscosity $\eta^s$, which is not parallel but perpendicular to the direction of the rotation. This dissipationless rotation response is a unique signature of parity-violating phase of a system with spinors degrees of freedom.

In Chern insulators, the spinor and orbital degrees of freedom are  locked to each other. In the case of a Dirac (weyl)  fermion, the spinor polarization is locked with the direction of propagation of the state (the momentum). In our case, the spinor polarization is  locked instead with quadrupole moment of the momentum of the state. In this way, a rotation in spinor space  induces a momentum current and vice versa. 

A consequence of these observations is that there must be a relation between the Hall viscosity and Hall torque viscosity. 
To see what the relation is let us compare the stress tensor with the spinor torque. Let us compute the rate of change of the action under an infinitesimal change of the frame fields, parametrized by $e_1$ and $e_2$ respectively (defined in Eq.\eqref{eq:frame-change}), and compare that with the torque. We obtain
\begin{align}
T_{ij}+T_{ji}=&\frac{\delta S}{\delta e_1}=-\bar \psi (2 \partial_1\partial_2 \gamma_1+(\partial_1^2-\partial_2^2)\gamma_2) \psi\nonumber\\
T_{ii}-T_{jj}=&\frac{\delta S}{\delta e_2}=-\bar \psi (-2 \partial_1\partial_2 \gamma_2+(\partial_1^2-\partial_2^2)\gamma_1) \psi 
\end{align}
and
\begin{align}
S_1=&\frac{\delta S}{\delta \theta_1}=\bar \psi ((\partial_1^2-\partial_2^2)+m \gamma_1) \psi \nonumber\\
S_2=&\frac{\delta S}{\delta \theta_2}=\bar \psi (2\partial_1 \partial_2+m \gamma_2) \psi 
\end{align}
After some simple algebra, it is easy to check the equivalence between spin rotation torque and the stress tensor,
\begin{align}
T_{ij}+T_{ji}=&\; -(S_1 \gamma_2+S_2 \gamma_1)\nonumber\\
T_{ii}-T_{jj}=&\; -(S_1 \gamma_1-S_2 \gamma_2)
\end{align}
As a result, if we subtract the antisymmetric parts from both the stress tensor correlator and of the torque correlator, we obtain
\begin{align}
- \left< \frac{\delta^2 S}{\delta \theta_1 \delta \theta_2} \right> + \left< \frac{\delta^2 S}{\delta M_1 \delta M_2} \right> = \left< \frac{\delta^2 S}{\delta e_1 \delta e_2}\right>
\label{eq:viscosity}
\end{align}
This identity implies the following linear relation between Hall viscosity $\eta$, the Hall torque viscosity $\eta^s$, and the Berry phase $\chi$ coefficient in our effective theory,
\begin{align}
-\eta^s+4\chi=\eta
\label{eq:relation}
\end{align}
Thus, the Berry phase term that was obtained from the effective theory for the nematic order parameter field measures the difference of Hall viscosity and Hall torque viscosity. We should note that the expressions for $\chi$, $\eta$ and $\eta_s$ given, respectively, in Eqs. \eqref{eq:chi}, \eqref{eq:eta0} and \eqref{eq:torque-viscosity},
obey this relation exactly.

The validity of these results are not restricted to the particular Chern insulator we studied here. The Hall torque viscosity is a universal property in all kinds of QAH phases. In systems  in which the fermions arise from of several orbitals,  the fermion operator in the effective action is a multi-component spinor. Suppose that the system has a non-vanishing Chern number, and hence that it is in a QAH state. If we rotate the spinor frame, the torque viscosity tensor, which is the linear response coefficient between torque and the angular velocity of the spinor rotation, must always include an antisymmetric part resulting from the parity violation in the fermion system. 

As an example, let us choose the case of a Dirac (Weyl) fermion.
Suppose we rotate the spinor frame in a similar way as in Eq.\eqref{eq:spinor-rotation}. After this rotation which, again is not a symmetry transformation, the Lagrangian changes to
\begin{align}
\mathcal{L}=\;
&\bar \psi'(x) (i \gamma_0 p_0- \gamma_1p_1- \gamma_2 p_2-m) \psi'(x)\nonumber\\
+\; &\bar \psi'(x)(\theta_1 p_1+\theta_2 p_2+m\theta_1 \gamma_1 + m\theta_2 \gamma_2) \psi'(x)
\end{align}
In the case of a Dirac (weyl) fermion the rotation metric couples both with the current and momentum. If we integrate-out the fermion, we would also get a Hall torque viscosity term
\begin{align}
\mathcal{L}(\theta)=
\frac{(-m \Lambda+4m^2)}{8\pi}  \epsilon^{ij} \theta_i \partial_0 \theta_j+\ldots
\end{align}
For a Dirac fermion, the spin is locked with linear momentum. Therefore, the equivalence between a spinor rotation and momentum current is expected and, hence, there is a similar relation between Hall viscosity and Hall torque viscosity.

\section{Conclusions} 
\label{sec:conclusions}

In this paper we presented a theory of the Mott quantum anomalous Hall state in the vicinity of its transition to a nematic QAH state. Our theory was developed in the context of a theory of spinless fermions which, at the free fermion level has a quadratic band crossing. A main result of this work is the effective field theory of Sections \ref{sec:1/N} and \ref{sec:nematic-effective-action} in which we derived the effective action for the hydrodynamic gauge fields $a_\mu$ and $b_\mu$ (which represent the charge currents) and 
the nematic order parameter field ${\bm M}$. The gauge theory sector is dominated by two topological terms, the BF term and the Chern-Simons term. The effective action of the nematic fields was found to contain a Berry phase term whose parity and time-reversal odd coefficient $\chi$ controls the dynamics. In particular the effective dynamical exponent of the nematic fields is $z=2$, consistent with the results of Maciejko {\it et al} developed in the context of the fractional quantum Hall states.\cite{Maciejko-2013} We also found that the nematic fields couple to the gauge field $a_\mu$ as a spatial metric. Our results clarify the role of geometric degrees of freedom in systems that exhibit the quantum Hall effect. We expect that these results should also apply to the case of the fractional quantum Hall effect and we will discuss these results elsewhere.\cite{You-2013}

In this work we considered the transition from the QAH phase to a nematic QAH phase (which is a continuous transition). It is is also possible to instead consider different regime of coupling constants in which  the leading instability from the QBC is to a  nematic semi-metal followed by a first order transition to the nematic QAH.\cite{kai} However in this case the theory that we presented here does not strictly apply since the transition would now be first order. Nevertheless the structure of our main results will still hold. A direct instability from the free QBC system to a nematic QAH phase does not seem to occur naturally.

In Section \ref{sec:torque-viscosity} we investigated the relation between the coefficient $\chi$ of the Berry phase of the nematic fields and the Hall viscosity $\eta$ of the spinors, which measures the transverse response to a local change of the spinor frame. Here we found that the complete picture requires the introduction of the concept of the torque Hall viscosity $\eta_s$, which is related to the fact that for s system of spinors a deformation of the underlying space requires the introduction of a spin connection. This effect is associated with the kinematics of spinors. Although it is always present multi-component fermionic systems, it takes a different form for Dirac fermions and in  this model with a quadratic band crossing (with unit Chern number). In particular we found that these three coefficients obey a universal linear relation given in Eq.\eqref{eq:relation}. Nevertheless these features are generic properties.

Our results are of interest in several systems accessible to experiment. One such system is bilayer graphene, which has two (almost exact) quadratic band crossings in the Brillouin zone. They are almost exact in that their quadratic band crossing is not protected by symmetry. However it is ``protected'' by the chemistry (and physics) of the orbitals of carbon which renders their parity-even gaps extremely small (and negligible in practice). This is  a point that has been investigated at length in the literature.\cite{Nandkishore2010,Vafek2010,Lemonik2012} However in the case of bilayer graphene it is necessary to include the spin degrees of freedom (which we suppressed here). This leads to a more complex (and interesting!) phase diagram\cite{kai,Vafek2010,Vafek2010b} which deserves further exploration. 

In the transport experiments of Xia {\it et al.}\cite{Xia-2010} on the 2DEG in the first Landau level  a large nematic susceptibility is seen in the longitudinal resistivities at finite temperature with a weak in-plane magnetic field. The results presented elsewhere in this paper  predict a similar behavior for the longitudinal resistivity at finite temperature in the QAH-nematic phase.

Other systems of great interest for which these results may be relevant are the topological crystalline insulators.\cite{Fu2011} Systems of these type have surface states (protected by mirror symmetry) which to a good approximation are described (at the level of the band structure) by a low-energy Hamiltonian with two quadratic band crossings. In materials such as Pb$_{1-x}$Sn$_x$Se and Pb$_{1-x}$Sn$_x$Te,  these crossings  which are expected to occur at the $X$ points on the edges of the surface Brillouin zone have been seen in ARPES and STM experiments.\cite{Hsieh2012,Hasan2012,Story2012,Hasan2013,Yazdani2013} However each quadratic crossing is found to be split into a pair of gapless Dirac cones. Although there are materials-specific symmetry breaking effects  that can explain these findings,\cite{Fang2012} it is also possible that the splittings may be driven by correlation effects, as in the case of the nematic semimetal phase discussed in Ref.[\onlinecite{kai}]. Nevertheless it is possible that these materials (or a close relative of them) may also exhibit a spontaneous quantum anomalous Hall phase such as the one discussed here (based on the work of Ref.[\onlinecite{kai}]) and  that the physics that we discussed here in detail may apply there too. Other materials in which these ideas may be relevant are the pyrochlore iridates.\cite{Yang2010,Wan-2011,Witczak-Krempa2012}

One of the motivations of this work, as we stated above, was to explore the interplay between the topological sector of these systems and the more microscopic ``geometric'' degrees of freedom. This issue was raised originally in the context of the experiments of Xia {\it et al.} in fractional quantum Hall states  in the first Landau level of the 2DEG\cite{Xia-2010} and has motivated several important theoretical developments.\cite{chetan,Mulligan-2011,Haldane-2011,Maciejko-2013} 
Much of that work (see, e.g. Ref.[\onlinecite{haldanemetric}]) has focused on the role of geometric changes at the microscopic level (i.e. at the length scale of the magnetic length). However, as we showed in this paper these ``geometric'' degrees of freedom can be self-organized into nematic order parameter fields whose fluctuations may manifest at even long length scales and hence may trigger a quantum phase transition of a nematic topological phase. In a separate publication\cite{You-2013} we will show how the ideas presented here extend to the case of the 2DEG in the fractional quantum Hall regime.

\begin{acknowledgments}
We would like to thanks Kai Sun, Taylor Hughes, Rob Leigh, Joseph Maciejko, Mike Mulligan, Chetan Nayak, Steve Kivelson, Benjamin Hsu, Shivaji Sondhi, Chen Fang, Gil Cho and Matthew Gilbert for helpful discussions. Y-Z would like to thank Bo Yang for intuitive advice. This work is supported in part by the National Science Foundation through the grant DMR-1064319 at the University of Illinois.
\end{acknowledgments}

\appendix

\section{The calculation of the effective gauge theory}
\label{app:effective-gauge-theory}

\begin{figure}[hbt]
 \vspace{0.2cm}
\begin{center}
 \subfigure[]{ \includegraphics[width=0.3\textwidth]{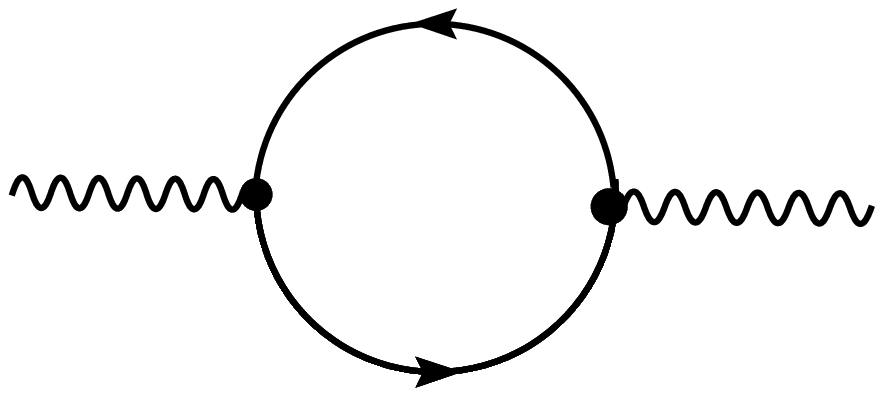} \label{fig:bubble}}
 \subfigure[]{\includegraphics[width=0.12\textwidth]{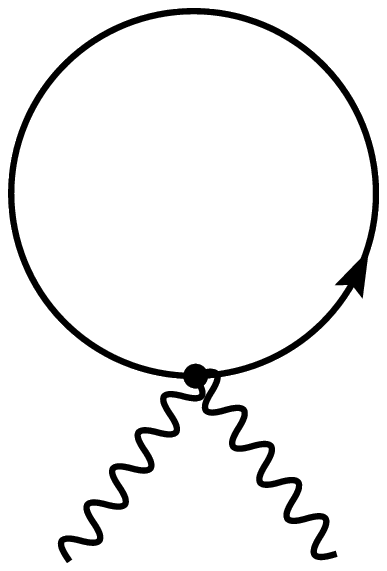} \label{fig:penguin}}
 \end{center}
\caption{One-loop self-energy diagrams for the hydrodynamic gauge field $a_\mu$.}
\end{figure}

To obtain the effective action of the gauge fields $S[a_\mu]$ we need to compute the one loop self-energy diagrams shown in Fig.\ref{fig:bubble} and Fig.\ref{fig:penguin}
Let  $G(p)$ be the fermion propagator of the quadratic band dispersion Chern insulator with mass $m$, i.e. in the isotropic QAH phase given in Eq.\eqref{eq:Feynman-propagator-pmu} with $\bm{M}=0$,
we can write the one-loop correction to the action $S^{(2)}[a_\mu]$ of the gauge fields in the standard form
\begin{equation}
S^{(2)}[a_\mu]=\frac{N}{2}\int \frac{d^3p}{(2\pi)^3}  a_{\mu}(-p) \Pi_{\mu \nu}(p)  a_{\nu}(p)
\end{equation}
$\Pi_{\mu \nu}(p)$ is the polarization operator which is the sum of two contributions: 
\begin{equation}
\Pi_{\mu \nu}(p)=\Pi^{(1)}_{\mu \nu}(p)+\Pi^{(2)}_{\mu \nu}(p)
\end{equation}
$\Pi_{\mu \nu}^{(1)}(p)$ is the diagram shown in Fig.\ref{fig:bubble} and is given by
\begin{align}
\Pi_{\mu \nu}^{(1)} &(p) =\nonumber\\
=i &\int \frac{d^3k}{(2\pi)^3} \textrm{tr}\Big[G_0(p+k) \mathcal{J}_{\mu}(2k+p) G_0(k) \mathcal{J}_{\nu}(2k+p)\Big]
\label{eq:pI1}
\end{align}
\begin{align}
\mathcal{J}_0(2k+p)=&\gamma_0  \\
\mathcal{J}_1(2k+p)=&  \gamma_1 (2k_1+p_1)+\gamma_2 (2k_2+p_2)\\
\mathcal{J}_2(2k+p)=&- \gamma_1 (2k_2+p_2)+ \gamma_2 (2k_1+p_1)
\end{align}
So the polarization tensor $\Pi_{\mu \nu}^{(1)} (p)$ has the expression,
\begin{widetext}
\begin{align}
\Pi_{\mu \nu}^{(1)} (p) 
=i \int \frac{d^3k}{(2\pi)^3} \textrm{tr}\Big[&
\frac{(p_0+k_0) \gamma_0-((p_1+k_1)^2-(p_2+k_2)^2)\gamma_1-2(p_1+k_1)(p_2+k_2)\gamma_2+m}{(p_0+k_0)^2-((p_1+k_1)^2+(p_2+k_2)^2)^2-m^2-i\epsilon}\\\nonumber
\times & \mathcal{J}_{\mu}(2k+p) \frac{k_0 \gamma_0-(k_1^2-k_2^2)\gamma_1-2k_1k_2\gamma_2+m}{k_0^2-(k_1^2+k_2^2)^2-m^2-i\epsilon} \mathcal{J}_{\nu}(2k+p)\Big]
\end{align}
\end{widetext}
As we only concern the long wave length behavior, we expand momentum p in $G_0(p+k)$ by order and only keeps $O(p^2)$.

\begin{figure}[hbt]
   \centering
   \subfigure[]{\includegraphics[width=0.22\textwidth]{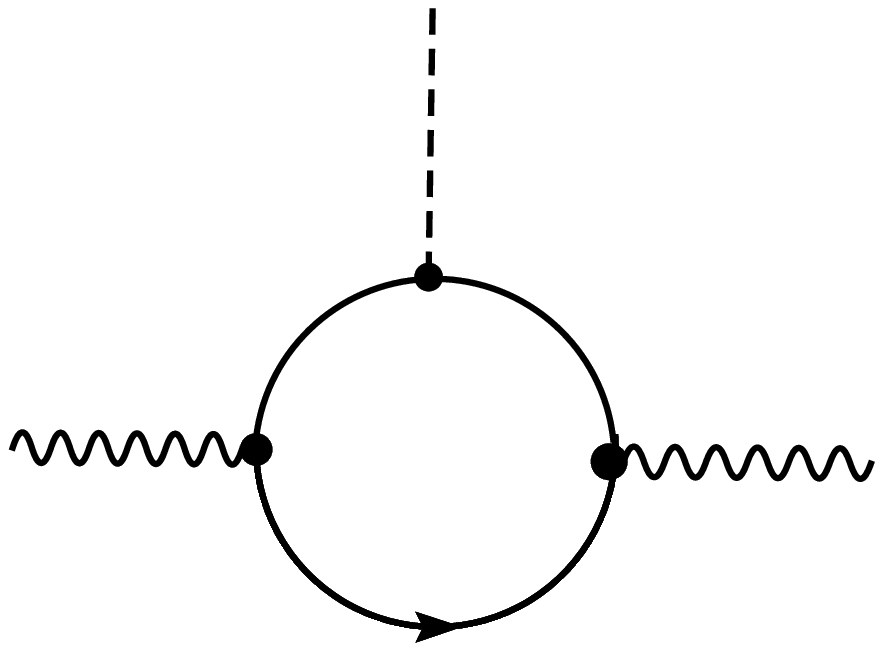}}
    \subfigure[]{\includegraphics[width=0.1\textwidth]{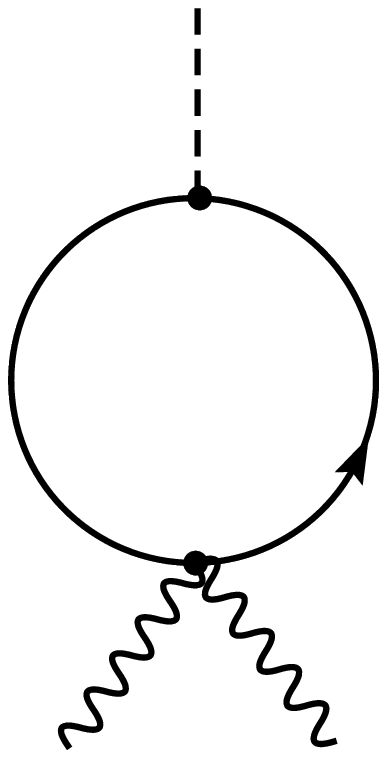}}
\caption{One-loop contributions to the vertex of the gauge field  $a_\mu$ and the nematic order parameter field ${\bm M}$.}
\label{fig:bubble+nematic}
\end{figure}

$\Pi_{\mu \nu}^{(2)}(p)$ is given by the diagram shown in Fig.\ref{fig:penguin} and is given by
\begin{equation}
\Pi_{\mu \nu}^{(2)}(p)=- i \int \frac{d^3k}{(2\pi)^3} \textrm{tr}[G_0(k) \bm{T}_{ij}] \\
\label{eq:Pi2}
\end{equation}
where $G_0(p)$ is the Feynman propagator of the isotropic QAH phase given in Eq.\eqref{eq:propagator}, and
\begin{align}
T_{11}=&\gamma_1,&
T_{22}=&-\gamma_1\nonumber\\
T_{12}=&\gamma_2,&
T_{21}=&\gamma_2
\label{eq:Tij}
\end{align}
In Eq.\eqref{eq:Pi2} the indices $\mu,\nu=i,j$ act only on the spatial components. Here we have to trace over all the matrix indices involved. Since there is either 
$\gamma_1$ or $\gamma_2$ in the expression for $T_{ij}$ (see Eq.\eqref{eq:Tij}), the only non-vanishing contribution to  the trace of
 \begin{equation}
 G_0(k)=\frac{k_0 \gamma_0-(k_1^2-k_2^2)\gamma_1-2k_1k_2\gamma_2+m}{k_0^2-(k_1^2+k_2^2)^2-m^2-i\epsilon}
 \end{equation}
should also include $\gamma_1$ or $\gamma_2$. However, these contributions have factors of  $k^2_1-k^2_2$ or $k_1 k_2$ in the numerator and hence cancel out we perform the after momentum integration. Thus we have 
\begin{equation}
\Pi_{\mu \nu}^{(2)}(p)=0
\end{equation}

The full one-loop polarization $\Pi_{\mu \nu}(p)$ is explicitly transverse. The resulting action $S[a_\mu]$ is gauge-invariant and is a sum of a parity-odd Chern-Simons term and a parity-even Maxwell term. The proof of gauge invariance is presented in Appendix \ref{app:proof}.

To obtain the leading coupling between nematic field and gauge field $S[a_\mu,\bm{M}]$, we need to calculate three-leg one-loop diagrams shown in Fig.\ref{fig:bubble+nematic} a and b.

\begin{align}
S[a_\mu,\bm{M}]=& \\ \nonumber
=\frac{N}{3}\int &\frac{d^3p}{(2\pi)^3} \Pi_{\mu \nu, i}(p_1,p_2)  a_{\mu}(-p_1-p_2) a_{\nu}(p_1) M_{i} (p_2)
\end{align}
There are two diagrams with non vanishing value,  so $\Pi_{\mu \nu ,i}$ are composed of two parts which are included in Fig.\ref{fig:bubble+nematic} a and b,
\begin{equation}
\Pi_{\mu \nu, i}(p_1,p_2) =\Pi_{\mu \nu,i}^{(1)}(p_1,p_2) +\Pi_{\mu \nu ,i} ^{(2)}(p_1) 
\end{equation}
where $p_1$ and $p_2$ are, respectively, the energy-momenta of the gauge field $a_\nu$ and of the nematic field $M_i$. Notice that $\Pi_{\mu \nu ,i} ^{(2)}(p_1)$ is only defined for $\mu, \nu=1,2$.

The one-loop three-legged diagram of  Fig.\ref{fig:bubble+nematic} a is 
\begin{widetext}
\begin{equation}
\Pi_{\mu \nu ,i}^{(1)}=-i\int \frac{d^3k}{(2\pi)^3}  \textrm{tr}[G_0 (k-p_2)\mathcal{J}_{\mu}(2k-p_2+p_1) 
G_0(k+p_1) \mathcal{J}_{\nu}(2k+p_1) G_0(k) \gamma_{i}]
\end{equation}
\end{widetext}
Note here greek symbol index labels the gauge field while latin symbol index labels the nematic field. The latin symbols only run for spatial index.
The one-loop diagram of Fig.\ref{fig:bubble+nematic}b has the expression (for $\mu=j$ and $\nu=k$)
\begin{equation}
\Pi_{j k ,i} ^{(2)}(p_2) =i\int \frac{d^3k}{(2\pi)^3} \; \textrm{tr}[G_0(-p_2+k) T_{jk} G_0(k) \gamma_{i}]
\end{equation}
where $T_{jk}$ is given in Eq.\eqref{eq:Tij}.

Here we expand the momentum $p_i$ by order and found the leading coupling term is the interplay between the nematic field and Maxwell term which is parity even. 
This is quite obvious. Since the gauge field enters quadratically in these diagrams, the leading gauge-invariant terms can only be the Chern-Simons term and Maxwell term. Since this theory is not Lorentz invariant, terms like $B \nabla \cdot E$  are allowed. We can ignore them as they are of higher order in derivatives than the Maxwell term. The Chern-Simons term is topological and as such it does not depend on the metric of the space-time. Thus, the only most relevant coupling should be the Maxwell term. This can also be seen from the polarization tensor. 

If we expand derivatives of nematic field $p_2$ in the polarization tensor by order, to the $O(1)$ order, we have
\begin{widetext}
\begin{align}
\Pi_{\mu \nu ,i}^{(1)}(p_1)&=-i\int \frac{d^3k}{(2\pi)^3}  \textrm{tr}\Big[G_0 (k)\mathcal{J}_{\mu}(2k+p_1) 
G_0(k+p_1) \mathcal{J}_{\nu}(2k+p_1) G_0(k) \gamma_{i}\Big] \\\nonumber
&=-i\int \frac{d^3k}{(2\pi)^3}  \textrm{tr}\Big[\mathcal{J}_{\mu}(2k+p_1) 
G_0(k+p_1) \mathcal{J}_{\nu}(2k+p_1) G_0(k) \gamma_{i} G_0 (k)\Big] 
\end{align}
\end{widetext}
If it is odd in $p_1$, the first terms in the products 
\begin{equation}
\mathcal{J}_{\mu}(2k+p_1) G_0(k+p_1) \mathcal{J}_{\nu}(2k+p_1) G_0(k)
\end{equation}
 being even and symmetric in the momentum $k$, should include a Levi-Civita tensor. In this sense, to obtain a non-vanishing value after trace, the $\gamma_{i} G_0 (k)$ term should not contribute any Gamma matrix. As a result, it would involve with $k_1^2-k_2^2$ which make the whole polarization tensor vanish after integration.

Upon expanding in derivatives of the nematic field $p_2$ to the $O(p_2)$ order, we have,
\begin{widetext}
\begin{align}
\Pi_{\mu \nu ,i}^{(1)}=&-i\int \frac{d^3k}{(2\pi)^3}  \textrm{tr}[G_0 (k)\mathcal{J}_{\mu}(-p_2) 
G_0(k+p_1) \mathcal{J}_{\nu}(2k+p_1) G_0(k) \gamma_{i}] \nonumber \\
&-i\int \frac{d^3k}{(2\pi)^3}  \textrm{tr}[\frac{F(p_2,k)}{k_0^2-(k_1^2+k_2^2)^2-m^2-i\epsilon} \mathcal{J}_{\mu}(2k+p_1) 
G_0(k+p_1) \mathcal{J}_{\nu}(2k+p_1) G_0(k) \gamma_{i}]
\label{eq:Pimunui1}
\end{align}
\end{widetext}
Here $F(p_2,k)$ is a function which is linear in $p_2$ and odd in $k$.

If it is odd in $p_1$, the second term of Eq.\eqref{eq:Pimunui1}
\begin{equation}
\mathcal{J}_{\mu}(2k+p_1) G_0(k+p_1) \mathcal{J}_{\nu}(2k+p_1) G_0(k)
\end{equation}
 includes a Levi-civita tensor and is even and symmetric in $k$. However, $F(p_2,k)$ is odd in $k$ and the integral vanishes.
For the first term of Eq.\eqref{eq:Pimunui1}, if $\mu=0$, $\mathcal{J}_{0}(-p_2) $ does not depend on $p_2$, this term is still of zeroth-order in $p_2$, and vanishes as we showed before. Otherwise, if it is odd in $p_1$, it is also odd in $k$ and the integral vanishes. Thus, to lowest order, there is no parity-odd coupling between the nematic field ${\bm M}$ and the gauge field $a_\mu$.

\section{The calculation of the effective nematic action}
\label{app:effective-action-nematic}

\begin{figure}[hbt]
   \centering
  \includegraphics[width=0.4\textwidth]{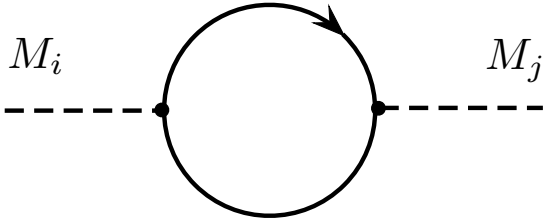}
\caption{One-loop self energy diagram for the nematic order parameter field.}
  \label{fig:nematic-self-energy}
\end{figure}

The only one-loop diagram that contributes is the  self-energy of the effective field theory of the nematic order parameter is  shown in Fig.\ref{fig:nematic-self-energy} and it is given by the expression
\begin{widetext}
\begin{align}
S_{\rm eff}(\bm M)&=-\frac {N}{2}\int d^3x d^3y \;  \textrm{tr} \left(G_{0}(x-y) {\bm \gamma} \cdot {\bm M}(y) G_{0}(y-x) {\bm \gamma} \cdot  {\bm M}(x)\right) -\int d^3x \; \frac {N}{2g} {\bm M}^2(x)\nonumber  \\
&= -\frac {N}{2} \int \frac{d^3 p}{(2\pi)^3} M_{i}(-p)\Gamma_{ij}(p) M_{j}(p) -\frac{N}{2g} \int \frac{d^3p}{(2\pi)^3} |{\bm M}(p)|^2
\end{align}
where $\Gamma_{ij}(p)$ is the one-loop kernel 
\begin{align}
\Gamma_{ij}(p) =&\int \frac{d^3 k}{(2\pi)^3} \textrm{tr}(\gamma_{i} G_{0}(p+k)\gamma_{j} G_{0}(k)) \\\nonumber
=&\epsilon^{ij} p_0 \int \frac{d^3k}{(2\pi)^3} \frac{m}{(k_0^2-(k_1^2+k_2^2)^2-m^2-i\epsilon)^2}
+\delta_{ij}\int \frac{d^3k}{(2\pi)^3}  \frac{m^2-k_0^2}{(k_0^2-(k_1^2+k_2^2)^2-m^2-i\epsilon)^2} +O(p^2)
\end{align}
\end{widetext}
The first term is odd in the frequency $p_0$ and contributes to the Berry phase term. The second term, which is even in the frequency $p_0$, contributes to the mass term of the nematic order parameter field and thus contains the information of the critical coupling constant for the quantum phase transition to the nematic phase.

\section{Short proof on gauge invariance of the polarization tensor}
\label{app:proof}

To verify the gauge invariance of the effective field theory we sketch here a proof on the gauge invariance of the polarization tensor. For the one-loop gauge field self-energy diagrams shown in Figs.\ref{fig:bubble}  and \ref{fig:penguin}. For gauge invariance to hold the polarization tensor should obey the transversality (conservation) condition
\begin{align} 
\Pi_{\mu \nu} p^{\nu}=0
\end{align}

Since the theory we start with is not Lorentz-invariant, the polarization tensor here decomposes into two parts, one of which, called $\Pi_{\mu \nu}^{(1)}$,  comes from the linear terms of the gauge field $a_\mu$ of the  Lagrangian, while $\Pi_{\mu \nu}^{(2)}$ comes from the terms which are quadratic in this gauge field,
\begin{align}
\Pi_{\mu \nu} =\Pi_{\mu \nu}^{(1)} +\Pi_{\mu \nu}^{(2)}  
\end{align}
We already showed  in appendix A that the second piece vanishes, $\Pi_{\mu \nu}^{(2)}=0.$
Hence, we only have to prove that
\begin{equation} 
\Pi_{\mu \nu}^{(1)} p^{\nu}=0
\end{equation}
Explicitly the left-hand-side of this equation is equal to
\begin{align}
\Pi_{\mu \nu}&p^{\nu}=\nonumber\\
=&\; \textrm{tr}\Big[ \int \frac{d^3k}{(2\pi)^3}  G_0 (p+k) \mathcal{J}_{\mu}(2k+p) G_0 (k) \mathcal{J}_{\nu}(2k+p) p_{\nu}\Big] \nonumber\\
&
\label{eq:Pimunu-transversality1}
\end{align}
Using the following Ward identity (whose validity is elementary to check) 
\begin{align} 
\mathcal{J}_{\nu}(2k+p) p^{\nu} =&\; p_0 \gamma_0 +(p_1^2-p_1^2+2p_1k_1-2p_2k_2) \gamma_1 
\nonumber\\
&+ 2(p_1p_2+p_1k_2+p_2k_1)\gamma_2\nonumber\\
=&\; G_0^{-1} (p+k) -G_0^{-1}(k)
\end{align}
we can write Eq.\eqref{eq:Pimunu-transversality1} in the form
\begin{widetext}
\begin{align} 
\Pi_{\mu \nu}p^{\nu} =&=\textrm{tr}\Big[\int \frac{d^3k}{(2\pi)^3} G_0 (p+k) \mathcal{J}_{\mu}(2k+p) G_0 (k) (G_0^{-1} (p+k) -G_0^{-1}(k) )\Big] \nonumber\\
&=\textrm{tr}\Big[\int \frac{d^3k}{(2\pi)^3} G_0 (p+k) \mathcal{J}_{\mu}(2k+p) \Big] 
-\textrm{tr}\Big[\int \frac{d^3k}{(2\pi)^3} G_0 (k) \mathcal{J}_{\mu}(2k+p) \Big]
\end{align}
For $\mu=0$, $\mathcal{J}_{0}=\gamma_0$,  we find
\begin{align} 
\textrm{tr}\Big[\int \frac{d^3k}{(2\pi)^3}  G_0 (p+k) \gamma_{0}\Big]-\textrm{tr}\Big[\int \frac{d^3k}{(2\pi)^3}  G_0 (k) \gamma_{0}\Big]=0
\end{align}
and for $\mu=1$, we get
\begin{align} 
\textrm{tr}\Big[\int \frac{d^3k}{(2\pi)^3}  G_0 (p+k) \mathcal{J}_{\mu}(2k+p)\Big]-\textrm{tr}\Big[\int \frac{d^3k}{(2\pi)^3}  G_0 (k) \mathcal{J}_{\mu}(2k+p)\Big]
&=-\textrm{tr}\Big[\int \frac{d^3k}{(2\pi)^3}  G_0 (p+k) (p_x \gamma_1+p_y \gamma_2) \Big]\nonumber\\
&=-\textrm{tr}\Big[\int \frac{d^3k}{(2\pi)^3}  G_0 (k) (p_x \gamma_1+p_y \gamma_2) \Big]
=0
\end{align}
Similarly,
for $\mu=2$ we also get
\begin{align} 
\textrm{tr}\Big[\int \frac{d^3k}{(2\pi)^3}  G_0 (p+k) \mathcal{J}_{\mu}(2k+p)\Big]
-\textrm{tr}\Big[\int \frac{d^3k}{(2\pi)^3}  G_0 (k) \mathcal{J}_{\mu}(2k+p)\Big]
&=-\textrm{tr}\Big[\int \frac{d^3k}{(2\pi)^3}  G_0 (p+k) (-p_y \gamma_1+p_x \gamma_2) \Big]\nonumber\\
&=-\textrm{tr}\Big[\int \frac{d^3k}{(2\pi)^3}  G_0 (k) (-p_y \gamma_1+p_x \gamma_2) \Big]
=0
\end{align}
\end{widetext}
Thus the polarization tensor of the gauge field $a_\mu$, the one-loop diagram of Fig.\ref{fig:bubble}, is transverse and, hence, the action of $a_\mu$ is gauge invariant.

We now turn to the gauge invariance of the coupling between the gauge field $a_\mu$ and the nematic order parameter field ${\bm M}$. The lowest order contribution to this coupling in the $1/N$ expansion is given by the Feynman diagrams shown in Fig.\ref{fig:bubble+nematic} a and Fig.\ref{fig:bubble+nematic} b. These diagrams contribute  to the effective action in the form
\begin{align}
&S[a_\mu,\bm{M}]\\\nonumber
&=\frac{N}{3}\int \frac{d^3p}{(2\pi)^3} \Pi_{\mu \nu i}(p_1,p_2)  a_{\mu}(-p_1-p_2) a_{\nu}(p_1) M_{i}(p_2)
\end{align}
Invariance under a gauge transformation $a_{\mu}+\partial_{\mu} \theta$ requires that this new polarization tensor, $\Pi_{\mu \nu i}(p_1,p_2)$,  should obey the following rule
\begin{align}
&\Pi_{\mu \nu, i}^{1}(p_1,p_2) p_1^{\nu} a_{\mu}(-p_1-p_2) M_{i}(p_2) \nonumber\\
&+\Pi_{\tau \sigma ,j}^{1}(p_1,p_2) (-p_1^{\tau}-p_2^{\tau})   a_{\sigma}(p_1) M_{j}(p_2) \nonumber\\
&+\Pi_{\alpha \beta, k}^{2}(p_1,p_2) (-p_1^\alpha-p_2^\alpha)  T_{\alpha \beta} a_{\beta} (p_1) M_{k}(p_2) \nonumber\\
&+\Pi_{\alpha \beta ,k}^{2}(p_1,p_2) p_{1}^\beta   T_{\alpha \beta} a_{\alpha} (-p_{1}-p_{2}) M_{k}(p_2)=0
\end{align}
where $\Pi_{\mu \nu, i}^{1}(p_1,p_2)$ is given by  
\begin{widetext}
\begin{align} 
\Pi^{1}_{\mu \nu ,i}(p_1,p_2) p_1^{\nu}
&=\textrm{tr}\Big[\int \frac{d^3k}{(2\pi)^3}  G_0 (k-p_2) \mathcal{J}_{\mu}G_0 (k+p_1) \mathcal{J}_{\nu} p_{\nu} G_0 (k) \gamma_{i}\Big]
\nonumber\\
&=\textrm{tr}\Big[\int \frac{d^3k}{(2\pi)^3}  G_0 (k-p_2) \mathcal{J}_{\mu}G_0 (k+p_1) (G_0^{-1} (k+p_1)-G_0^{-1} (k))G_0 (k) \gamma_{i}\Big]\nonumber\\
&=\textrm{tr}\Big[\int \frac{d^3k}{(2\pi)^3}  G_0 (k-p_2) \mathcal{J}_{\mu}G_0 (k) \gamma_{i}\Big]
-\textrm{tr}\Big[\int \frac{d^3k}{(2\pi)^3}  G_0 (k-p_2) \mathcal{J}_{\mu} G_0 (k+p_1) \gamma_{i}\Big]
\end{align}

For $\mu=0$, $\mathcal{J}_0=\gamma_0$, we get
\begin{align} 
\Pi^{1}_{\mu \nu,i}(p_1,p_2) p_1^{\nu} 
&=\textrm{tr}\Big[\int \frac{d^3k}{(2\pi)^3}  G_0 (k-p_2) \gamma_0 G_0 (k) \gamma_{i}\Big]
-\textrm{tr}\Big[\int \frac{d^3k}{(2\pi)^3}  G_0 (k-p_2) \gamma_0 G_0 (k+p_1) \gamma_{i}\Big]\nonumber\\
&=\textrm{tr}\Big[\int \frac{d^3k}{(2\pi)^3}  G_0 (k-p_2) \gamma_0 G_0 (k) \gamma_{i}\Big] 
-\textrm{tr}\Big[\int \frac{d^3k}{(2\pi)^3}  G_0 (k-p_2) \gamma_0 G_0 (k+p_1) \gamma_{i}\Big]\nonumber\\
&=\textrm{tr}\Big[\int \frac{d^3k}{(2\pi)^3}  G_0 (k-p_2) \gamma_0 G_0 (k) \gamma_{i}\Big]
-\textrm{tr}\Big[\int \frac{d^3k}{(2\pi)^3}  G_0 (k-p_2-p_1) \gamma_0 G_0 (k) \gamma_{i}\Big]
\end{align}

Likewise, for $\sigma=0$, $\mathcal{J}_0=\gamma_0$, we obtain
\begin{align} 
\Pi^{1}_{\tau \sigma,j}(p_1,p_2) (-p_{1}^\tau-p_{2}^\tau )
&=\textrm{tr}\Big[\int \frac{d^3k}{(2\pi)^3}  G_0 (k-p_2) \mathcal{J}_{\tau} p_{\tau} G_0 (k+p_1) \mathcal{J}_{\sigma} G_0 (k) \gamma_{j}\Big]\nonumber\\
&=\textrm{tr}\Big[\int \frac{d^3k}{(2\pi)^3}  G_0 (k-p_2) (G_0^{-1} (k-p_2) 
-G_0^{-1} (k+p_1)) G_0 (k+p_1) \mathcal{J}_{\sigma} G_0 (k) \gamma_{j}\Big]\nonumber\\
&=\textrm{tr}\Big[\int \frac{d^3k}{(2\pi)^3}  G_0 (k+p_1) \mathcal{J}_{\nu} G_0 (k) \gamma_{j}\Big]
-\textrm{tr}\Big[\int \frac{d^3k}{(2\pi)^3}  G_0 (k-p_2) \mathcal{J}_{\sigma} G_0 (k) \gamma_{j}\Big]
\end{align}
Hence
\begin{equation} 
\Pi^{1}_{\tau \sigma ,j}(p_1,p_2) (-p_{1}^\tau-p_{2}^\tau )
=\textrm{tr}\Big[\int \frac{d^3k}{(2\pi)^3}  G_0 (k+p_1) \gamma_0 G_0 (k) \gamma_{j}\Big]
-\textrm{tr}\Big[\int \frac{d^3k}{(2\pi)^3}  G_0 (k-p_2) \gamma_0 G_0 (k) \gamma_{j}\Big]
\end{equation}
It is easy to check that for each $\nu=\tau$, $\mu=\sigma=0$,
\begin{equation} 
\Pi^{1}_{\mu \nu ,i}(p_1,p_2) p_1^{\nu} a_{\mu}(-p_1-p_2) M_{i}(p_2)
=-\Pi^{1}_{\tau \sigma ,j}(p_1,p_2) (-p_{1}^\tau-p_{2}^\tau )  a_{\sigma}(p_1) M_{j}(p_2).
\end{equation}

For $\mu=1$ we get
\begin{align} 
\Pi^{1}_{\mu \nu, i}(p_1,p_2) p_1^{\nu} 
&=\textrm{tr}\Big[\int \frac{d^3k}{(2\pi)^3}  G_0 (k-p_2) \mathcal{J}_1(2k-p_2+p_1) G_0 (k) \gamma_{i}\Big] \nonumber\\
    &-\textrm{tr}\Big[ \int \frac{d^3k}{(2\pi)^3}  G_0 (k-p_2)  \mathcal{J}_1(2k-p_2+p_1) G_0 (k+p_1) \gamma_{i}\Big]
\nonumber\\
&=\textrm{tr}\Big[\int \frac{d^3k}{(2\pi)^3}  G_0 (k-p_2) \mathcal{J}_1(2k-p_2+p_1) G_0 (k) \gamma_{i}\Big] 
\nonumber\\
 &  -\textrm{tr}\Big[\int \frac{d^3k}{(2\pi)^3}  G_0 (k-p_2-p_1)  \mathcal{J}_1(2k-p_2-p_1) G_0 (k) \gamma_{i}\Big]
\nonumber\\
&=\textrm{tr}\Big[\int \frac{d^3k}{(2\pi)^3}  G_0 (k-p_2) \mathcal{J}_1(2k-p_2) G_0 (k) \gamma_{i}\Big]
    +\textrm{tr}\Big[\int \frac{d^3k}{(2\pi)^3}  G_0 (k-p_2) \mathcal{J}_1(p_1) G_0 (k) \gamma_{i}\Big]
    \nonumber\\
& -\textrm{tr}\Big[\int \frac{d^3k}{(2\pi)^3}  G_0 (k-p_2-p_1)  \mathcal{J}_1(2k-p_2-p_1) G_0 (k) \gamma_{i}\Big]
\end{align}
and for $\sigma=1$,
\begin{align} 
\Pi^{1}_{\tau \sigma ,j}(p_1,p_2) (-p_{1}^\tau-p_{2}^\tau )
=& \:\textrm{tr}\Big[\int  \frac{d^3k}{(2\pi)^3}  G_0 (k+p_1) \mathcal{J}_1(2k+p_1) G_0 (k) \gamma_{j}] 
-\textrm{tr}\Big[\int  \frac{d^3k}{(2\pi)^3}  G_0 (k-p_2) \mathcal{J}_1(2k+p_1) G_0 (k) \gamma_{j} \Big]\nonumber\\
=& \: \textrm{tr}[ \int \frac{d^3k}{(2\pi)^3}  G_0 (k+p_1) \mathcal{J}_1(2k+p_1) G_0 (k) \gamma_{j} \Big] \nonumber\\
&-\textrm{tr}\Big[\int \frac{d^3k}{(2\pi)^3}  G_0 (k-p_2) \mathcal{J}_1(2k-p_2+p_2+p_1) G_0 (k) \gamma_{j} \Big]\nonumber\\
=& \; \textrm{tr}\Big[\int \frac{d^3k}{(2\pi)^3}  G_0 (k+p_1) \mathcal{J}_1(2k+p_1) G_0 (k) \gamma_{j} \Big]
-\textrm{tr}\Big[ \int \frac{d^3k}{(2\pi)^3}  G_0 (k-p_2) \mathcal{J}_1(2k-p_2) G_0 (k) \gamma_{j} \Big] \nonumber\\
&+\textrm{tr} \Big[ \int \frac{d^3k}{(2\pi)^3}  G_0 (k-p_2) \mathcal{J}_1(-p_2-p_1) G_0 (k) \gamma_{j} \Big]
\end{align}

After some algebra, it could be checked that the rest terms after a gauge transformation are,
\begin{align}
\Pi^{1}_{1 \nu, i}(p_1,p_2) p_1^{\nu} a_{1}(-p_1-p_2) M_{i}(p_2)
&+\Pi^{1}_{\tau 1 ,i}(p_1,p_2) (-p_{1}^\tau-p_{2}^\tau )  a_{1}(p_1) M_{i}(p_2) \nonumber\\
&=2 \; \textrm{tr} \Big[\int \frac{d^3k}{(2\pi)^3}  G_0 (k-p_2) 
 \mathcal{J}_1(-p_2-p_1) G_0 (k) \gamma_{i}\Big] a_{1}(p_1) M_{i}(p_2)
\end{align}

This contribution is cancelled by the ``tadpole+nematic'' diagram of Fig.\ref{fig:bubble+nematic} b. Indeed, up to a gauge transformation, the extra terms generated in this diagram are
\begin{align} 
\Pi_{\alpha 1 ,k}^{2}(p_1,p_2) (-p_{1}^{\alpha}-p_{2}^{\alpha})  T_{\alpha 1} a_1 (p_1) M_{k}(p_2)
&+\Pi_{1 \beta ,k}^{2}(p_1,p_2) p_{1}^{\beta }   T_{1 \beta } a_1 (-p_{1}-p_{2}) M_{k}(p_2)\nonumber \\
&=-2\; \textrm{tr} \Big[ \int \frac{d^3k}{(2\pi)^3}  G_0 (k-p) (\gamma_1 p'_x+\gamma_2 p'_y) G_0 (k)  \gamma_{k} \Big] a_{1}(-p-p' ) M_{k}(p)\nonumber\\
&=-2 \; \textrm{tr}\Big[\int \frac{d^3k}{(2\pi)^3}  G_0 (k-p) \mathcal{J}_1(p') G_0 (k)  \gamma_{k} \Big] a_{1}(-p-p' ) M_{k}(p)
\end{align}
which exactly cancels the offending terms.

In the case of the three-legged diagram which defines the tensor $\Pi^{1}_{2\nu, i}(p_1,p_2)$ we also obtain the same condition 
 for $\mu=\sigma=2$. The remaining terms, after a gauge transformation, are
\begin{align} 
\Pi^{1}_{2\nu, i}(p_1,p_2) p_1^{\nu} a_{2}(-p_1-p_2) M_{i}(p_2) 
&+\Pi^{1}_{\tau 2, i}(p_1,p_2) (-p_{1}^\tau-p_{2}^\tau )   a_{2}(p_1) M_{i}(p_2)\nonumber\\
&=2 \;\textrm{tr}[\int \frac{d^3k}{(2\pi)^3}  G_0 (k-p_2) 
 \mathcal{J}_{2}(-p_2-p_1) G_0 (k) \gamma_{i}] a_{2}(p_1) M_{i}(p_2)
\end{align}
This contribution is canceled by the extra terms in the tadpole+nematic diagram, Fig.\ref{fig:bubble+nematic} b,  after the gauge transformation
\begin{align} 
-2\; \textrm{tr}\Big[ \int \frac{d^3k}{(2\pi)^3}  G_0 (k-p) (\gamma_2 p'_x-\gamma_1 p'_y) G_0 (k)  \gamma_{k}\Big] a_{2}(p' )& M_{k}(p) 
=\nonumber\\
&=-2\; \textrm{tr}\Big[\int \frac{d^3k}{(2\pi)^3}  G_0 (k-p) \mathcal{J}_2(p') G_0 (k)  \gamma_{k} \Big] a_{2}(-p-p' ) M_{k}(p)
\end{align}
So the polarization tensor is transverse and the action is gauge-invariant (as it should be).

\end{widetext}


\begin{thebibliography}{90}
\expandafter\ifx\csname natexlab\endcsname\relax\def\natexlab#1{#1}\fi
\expandafter\ifx\csname bibnamefont\endcsname\relax
  \def\bibnamefont#1{#1}\fi
\expandafter\ifx\csname bibfnamefont\endcsname\relax
  \def\bibfnamefont#1{#1}\fi
\expandafter\ifx\csname citenamefont\endcsname\relax
  \def\citenamefont#1{#1}\fi
\expandafter\ifx\csname url\endcsname\relax
  \def\url#1{\texttt{#1}}\fi
\expandafter\ifx\csname urlprefix\endcsname\relax\def\urlprefix{URL }\fi
\providecommand{\bibinfo}[2]{#2}
\providecommand{\eprint}[2][]{\url{#2}}

\bibitem[{\citenamefont{{von Klitzing} et~al.}(1980)\citenamefont{{von
  Klitzing}, Dorda, and Pepper}}]{Klitzing-1980}
\bibinfo{author}{\bibfnamefont{K.}~\bibnamefont{{von Klitzing}}},
  \bibinfo{author}{\bibfnamefont{G.}~\bibnamefont{Dorda}}, \bibnamefont{and}
  \bibinfo{author}{\bibfnamefont{M.}~\bibnamefont{Pepper}},
  \bibinfo{journal}{Phys. Rev. Lett.} \textbf{\bibinfo{volume}{45}},
  \bibinfo{pages}{494} (\bibinfo{year}{1980}).

\bibitem[{\citenamefont{Tsui et~al.}(1982)\citenamefont{Tsui, Stormer, and
  Gossard}}]{Tsui-1982}
\bibinfo{author}{\bibfnamefont{D.~C.} \bibnamefont{Tsui}},
  \bibinfo{author}{\bibfnamefont{H.~L.} \bibnamefont{Stormer}},
  \bibnamefont{and} \bibinfo{author}{\bibfnamefont{A.~C.}
  \bibnamefont{Gossard}}, \bibinfo{journal}{Phys. Rev. Lett.}
  \textbf{\bibinfo{volume}{48}}, \bibinfo{pages}{1559} (\bibinfo{year}{1982}).

\bibitem[{\citenamefont{Laughlin}(1981)}]{Laughlin-1981}
\bibinfo{author}{\bibfnamefont{R.~B.} \bibnamefont{Laughlin}},
  \bibinfo{journal}{Phys. Rev. B} \textbf{\bibinfo{volume}{23}},
  \bibinfo{pages}{5632} (\bibinfo{year}{1981}).

\bibitem[{\citenamefont{Thouless et~al.}(1982)\citenamefont{Thouless, Kohmoto,
  Nightingale, and {den Nijs}}}]{Thouless-1982}
\bibinfo{author}{\bibfnamefont{D.~J.} \bibnamefont{Thouless}},
  \bibinfo{author}{\bibfnamefont{M.}~\bibnamefont{Kohmoto}},
  \bibinfo{author}{\bibfnamefont{M.~P.} \bibnamefont{Nightingale}},
  \bibnamefont{and} \bibinfo{author}{\bibfnamefont{M.}~\bibnamefont{{den
  Nijs}}}, \bibinfo{journal}{Phys. Rev. Lett.} \textbf{\bibinfo{volume}{49}},
  \bibinfo{pages}{405} (\bibinfo{year}{1982}).

\bibitem[{\citenamefont{Niu et~al.}(1985)\citenamefont{Niu, Thouless, and
  Wu}}]{Niu-1985}
\bibinfo{author}{\bibfnamefont{Q.}~\bibnamefont{Niu}},
  \bibinfo{author}{\bibfnamefont{D.~J.} \bibnamefont{Thouless}},
  \bibnamefont{and} \bibinfo{author}{\bibfnamefont{Y.-S.} \bibnamefont{Wu}},
  \bibinfo{journal}{Phys. Rev. B} \textbf{\bibinfo{volume}{31}},
  \bibinfo{pages}{3372} (\bibinfo{year}{1985}).

\bibitem[{\citenamefont{Laughlin}(1983)}]{Laughlin-1983}
\bibinfo{author}{\bibfnamefont{R.~B.} \bibnamefont{Laughlin}},
  \bibinfo{journal}{Phys. Rev. Lett.} \textbf{\bibinfo{volume}{50}},
  \bibinfo{pages}{1395} (\bibinfo{year}{1983}).

\bibitem[{\citenamefont{Haldane}(1983)}]{Haldane-1983}
\bibinfo{author}{\bibfnamefont{F.~D.~M.} \bibnamefont{Haldane}},
  \bibinfo{journal}{Phys. Rev. Lett.} \textbf{\bibinfo{volume}{51}},
  \bibinfo{pages}{605} (\bibinfo{year}{1983}).

\bibitem[{\citenamefont{Halperin}(1984)}]{Halperin-1984}
\bibinfo{author}{\bibfnamefont{B.~I.} \bibnamefont{Halperin}},
  \bibinfo{journal}{Phys. Rev. Lett.} \textbf{\bibinfo{volume}{52}},
  \bibinfo{pages}{1583} (\bibinfo{year}{1984}).

\bibitem[{\citenamefont{Wen and Niu}(1990)}]{Wen-1990}
\bibinfo{author}{\bibfnamefont{X.~G.} \bibnamefont{Wen}} \bibnamefont{and}
  \bibinfo{author}{\bibfnamefont{Q.}~\bibnamefont{Niu}},
  \bibinfo{journal}{Phys. Rev. B} \textbf{\bibinfo{volume}{41}},
  \bibinfo{pages}{9377} (\bibinfo{year}{1990}).

\bibitem[{\citenamefont{Zhang et~al.}(1989)\citenamefont{Zhang, Hansson, and
  Kivelson}}]{Zhang-1989}
\bibinfo{author}{\bibfnamefont{S.~C.} \bibnamefont{Zhang}},
  \bibinfo{author}{\bibfnamefont{T.~H.} \bibnamefont{Hansson}},
  \bibnamefont{and} \bibinfo{author}{\bibfnamefont{S.}~\bibnamefont{Kivelson}},
  \bibinfo{journal}{Phys. Rev. Lett.} \textbf{\bibinfo{volume}{62}},
  \bibinfo{pages}{82} (\bibinfo{year}{1989}).

\bibitem[{\citenamefont{L\'opez and Fradkin}(1991)}]{Lopez-1991}
\bibinfo{author}{\bibfnamefont{A.}~\bibnamefont{L\'opez}} \bibnamefont{and}
  \bibinfo{author}{\bibfnamefont{E.}~\bibnamefont{Fradkin}},
  \bibinfo{journal}{Phys. Rev. B} \textbf{\bibinfo{volume}{44}},
  \bibinfo{pages}{5246} (\bibinfo{year}{1991}).

\bibitem[{\citenamefont{Fr{\"o}hlich and Kerler}(1991)}]{Frohlich1991}
\bibinfo{author}{\bibfnamefont{J.}~\bibnamefont{Fr{\"o}hlich}}
  \bibnamefont{and} \bibinfo{author}{\bibfnamefont{T.}~\bibnamefont{Kerler}},
  \bibinfo{journal}{Nucl. Phys. B} \textbf{\bibinfo{volume}{354}},
  \bibinfo{pages}{369} (\bibinfo{year}{1991}).

\bibitem[{\citenamefont{Wen and Zee}(1992{\natexlab{a}})}]{Wen-1992}
\bibinfo{author}{\bibfnamefont{X.-G.} \bibnamefont{Wen}} \bibnamefont{and}
  \bibinfo{author}{\bibfnamefont{A.}~\bibnamefont{Zee}},
  \bibinfo{journal}{Phys. Rev. B} \textbf{\bibinfo{volume}{46}},
  \bibinfo{pages}{2290} (\bibinfo{year}{1992}{\natexlab{a}}).

\bibitem[{\citenamefont{Wen}(1995{\natexlab{a}})}]{Wen-1995}
\bibinfo{author}{\bibfnamefont{X.~G.} \bibnamefont{Wen}},
  \bibinfo{journal}{Adv. Phys.} \textbf{\bibinfo{volume}{44}},
  \bibinfo{pages}{405} (\bibinfo{year}{1995}{\natexlab{a}}).

\bibitem[{\citenamefont{Jiang et~al.}(2012)\citenamefont{Jiang, Yao, and
  Balents}}]{Jiang-2011}
\bibinfo{author}{\bibfnamefont{H.~C.} \bibnamefont{Jiang}},
  \bibinfo{author}{\bibfnamefont{H.}~\bibnamefont{Yao}}, \bibnamefont{and}
  \bibinfo{author}{\bibfnamefont{L.}~\bibnamefont{Balents}},
  \bibinfo{journal}{Phys. Rev. B} \textbf{\bibinfo{volume}{86}},
  \bibinfo{pages}{024424} (\bibinfo{year}{2012}).

\bibitem[{\citenamefont{Rokhsar and Kivelson}(1988)}]{Rokhsar-1988}
\bibinfo{author}{\bibfnamefont{D.}~\bibnamefont{Rokhsar}} \bibnamefont{and}
  \bibinfo{author}{\bibfnamefont{S.~A.} \bibnamefont{Kivelson}},
  \bibinfo{journal}{Phys. Rev. Lett.} \textbf{\bibinfo{volume}{61}},
  \bibinfo{pages}{2376} (\bibinfo{year}{1988}).

\bibitem[{\citenamefont{Moessner and Sondhi}(2001)}]{Moessner-2001}
\bibinfo{author}{\bibfnamefont{R.}~\bibnamefont{Moessner}} \bibnamefont{and}
  \bibinfo{author}{\bibfnamefont{S.~L.} \bibnamefont{Sondhi}},
  \bibinfo{journal}{Phys. Rev. Lett.} \textbf{\bibinfo{volume}{86}},
  \bibinfo{pages}{1881} (\bibinfo{year}{2001}).

\bibitem[{\citenamefont{Bernevig et~al.}(2006)\citenamefont{Bernevig, Hughes,
  and Zhang}}]{Bernevig-2006b}
\bibinfo{author}{\bibfnamefont{B.~A.} \bibnamefont{Bernevig}},
  \bibinfo{author}{\bibfnamefont{T.~L.} \bibnamefont{Hughes}},
  \bibnamefont{and} \bibinfo{author}{\bibfnamefont{S.~C.} \bibnamefont{Zhang}},
  \bibinfo{journal}{Science} \textbf{\bibinfo{volume}{314}},
  \bibinfo{pages}{1757} (\bibinfo{year}{2006}).

\bibitem[{\citenamefont{K\"{o}nig et~al.}(2007)\citenamefont{K\"{o}nig,
  Steffen, Br{\"u}ne, Roth, Buhmann, Molenkamp, Qi, and Zhang}}]{Konig-2007}
\bibinfo{author}{\bibfnamefont{M.}~\bibnamefont{K\"{o}nig}},
  \bibinfo{author}{\bibfnamefont{S.}~\bibnamefont{Steffen}},
  \bibinfo{author}{\bibfnamefont{C.}~\bibnamefont{Br{\"u}ne}},
  \bibinfo{author}{\bibfnamefont{A.}~\bibnamefont{Roth}},
  \bibinfo{author}{\bibfnamefont{H.}~\bibnamefont{Buhmann}},
  \bibinfo{author}{\bibfnamefont{L.~W.} \bibnamefont{Molenkamp}},
  \bibinfo{author}{\bibfnamefont{X.~L.} \bibnamefont{Qi}}, \bibnamefont{and}
  \bibinfo{author}{\bibfnamefont{S.~C.} \bibnamefont{Zhang}},
  \bibinfo{journal}{Science} \textbf{\bibinfo{volume}{318}},
  \bibinfo{pages}{766} (\bibinfo{year}{2007}).

\bibitem[{\citenamefont{Fu et~al.}(2007)\citenamefont{Fu, Kane, and
  Mele}}]{Fu-2007b}
\bibinfo{author}{\bibfnamefont{L.}~\bibnamefont{Fu}},
  \bibinfo{author}{\bibfnamefont{C.~L.} \bibnamefont{Kane}}, \bibnamefont{and}
  \bibinfo{author}{\bibfnamefont{E.~J.} \bibnamefont{Mele}},
  \bibinfo{journal}{Phys. Rev. Lett.} \textbf{\bibinfo{volume}{98}},
  \bibinfo{pages}{106803} (\bibinfo{year}{2007}).

\bibitem[{\citenamefont{Hasan and Kane}(2010)}]{Hasan-2010}
\bibinfo{author}{\bibfnamefont{M.~Z.} \bibnamefont{Hasan}} \bibnamefont{and}
  \bibinfo{author}{\bibfnamefont{C.~L.} \bibnamefont{Kane}},
  \bibinfo{journal}{Rev. Mod. Phys.} \textbf{\bibinfo{volume}{82}},
  \bibinfo{pages}{3045} (\bibinfo{year}{2010}).

\bibitem[{\citenamefont{Hasan and Moore}(2011)}]{Hasan-2011}
\bibinfo{author}{\bibfnamefont{M.~Z.} \bibnamefont{Hasan}} \bibnamefont{and}
  \bibinfo{author}{\bibfnamefont{J.~E.} \bibnamefont{Moore}},
  \bibinfo{journal}{Annual Reviews of Condensed Matter Physics}
  \textbf{\bibinfo{volume}{2}}, \bibinfo{pages}{55} (\bibinfo{year}{2011}).

\bibitem[{\citenamefont{Haldane}(1988)}]{qah}
\bibinfo{author}{\bibfnamefont{F.~D.~M.} \bibnamefont{Haldane}},
  \bibinfo{journal}{Phys. Rev. Lett.} \textbf{\bibinfo{volume}{61}},
  \bibinfo{pages}{2015} (\bibinfo{year}{1988}).

\bibitem[{\citenamefont{Neupert et~al.}(2011)\citenamefont{Neupert, Santos,
  Chamon, and Mudry}}]{Neupert-2011}
\bibinfo{author}{\bibfnamefont{T.}~\bibnamefont{Neupert}},
  \bibinfo{author}{\bibfnamefont{L.}~\bibnamefont{Santos}},
  \bibinfo{author}{\bibfnamefont{C.}~\bibnamefont{Chamon}}, \bibnamefont{and}
  \bibinfo{author}{\bibfnamefont{C.}~\bibnamefont{Mudry}},
  \bibinfo{journal}{Phys. Rev. Lett.} \textbf{\bibinfo{volume}{106}},
  \bibinfo{pages}{236804} (\bibinfo{year}{2011}).

\bibitem[{\citenamefont{Sheng et~al.}(2011)\citenamefont{Sheng, Gu, Sun, and
  Sheng}}]{Sheng-2011}
\bibinfo{author}{\bibfnamefont{D.~N.} \bibnamefont{Sheng}},
  \bibinfo{author}{\bibfnamefont{Z.~C.} \bibnamefont{Gu}},
  \bibinfo{author}{\bibfnamefont{K.}~\bibnamefont{Sun}}, \bibnamefont{and}
  \bibinfo{author}{\bibfnamefont{L.}~\bibnamefont{Sheng}},
  \bibinfo{journal}{Nature Communications} \textbf{\bibinfo{volume}{2}},
  \bibinfo{pages}{389} (\bibinfo{year}{2011}).

\bibitem[{\citenamefont{Regnault and Bernevig}(2011)}]{Regnault-2011}
\bibinfo{author}{\bibfnamefont{N.}~\bibnamefont{Regnault}} \bibnamefont{and}
  \bibinfo{author}{\bibfnamefont{B.~A.} \bibnamefont{Bernevig}},
  \bibinfo{journal}{Phys. Rev. X} \textbf{\bibinfo{volume}{1}},
  \bibinfo{pages}{021014} (\bibinfo{year}{2011}).

\bibitem[{\citenamefont{Cincio and Vidal}(2013)}]{Cinicio-2013}
\bibinfo{author}{\bibfnamefont{L.}~\bibnamefont{Cincio}} \bibnamefont{and}
  \bibinfo{author}{\bibfnamefont{G.}~\bibnamefont{Vidal}},
  \bibinfo{journal}{Phys. Rev. Lett.} \textbf{\bibinfo{volume}{110}},
  \bibinfo{pages}{067208} (\bibinfo{year}{2013}).

\bibitem[{\citenamefont{Sondhi et~al.}(1993)\citenamefont{Sondhi, Karlhede,
  Kivelson, and Rezayi}}]{Sondhi-1993}
\bibinfo{author}{\bibfnamefont{S.~L.} \bibnamefont{Sondhi}},
  \bibinfo{author}{\bibfnamefont{A.}~\bibnamefont{Karlhede}},
  \bibinfo{author}{\bibfnamefont{S.~A.} \bibnamefont{Kivelson}},
  \bibnamefont{and} \bibinfo{author}{\bibfnamefont{E.~H.}
  \bibnamefont{Rezayi}}, \bibinfo{journal}{Phys. Rev. B}
  \textbf{\bibinfo{volume}{47}}, \bibinfo{pages}{16419} (\bibinfo{year}{1993}).

\bibitem[{\citenamefont{Ho}(1994)}]{Ho-1994}
\bibinfo{author}{\bibfnamefont{T.-L.} \bibnamefont{Ho}},
  \bibinfo{journal}{Phys. Rev. Lett.} \textbf{\bibinfo{volume}{73}},
  \bibinfo{pages}{874} (\bibinfo{year}{1994}).

\bibitem[{\citenamefont{Shkolnikov et~al.}(2005)\citenamefont{Shkolnikov,
  Misra, Bishop, De~Poortere, and Shayegan}}]{Shkolnikov-2005}
\bibinfo{author}{\bibfnamefont{Y.~P.} \bibnamefont{Shkolnikov}},
  \bibinfo{author}{\bibfnamefont{S.}~\bibnamefont{Misra}},
  \bibinfo{author}{\bibfnamefont{N.~C.} \bibnamefont{Bishop}},
  \bibinfo{author}{\bibfnamefont{E.~P.} \bibnamefont{De~Poortere}},
  \bibnamefont{and} \bibinfo{author}{\bibfnamefont{M.}~\bibnamefont{Shayegan}},
  \bibinfo{journal}{Phys. Rev. Lett.} \textbf{\bibinfo{volume}{95}},
  \bibinfo{pages}{066809} (\bibinfo{year}{2005}).

\bibitem[{\citenamefont{Abanin et~al.}(2010)\citenamefont{Abanin, Parameswaran,
  Kivelson, and Sondhi}}]{Abanin-2010}
\bibinfo{author}{\bibfnamefont{D.~A.} \bibnamefont{Abanin}},
  \bibinfo{author}{\bibfnamefont{S.~A.} \bibnamefont{Parameswaran}},
  \bibinfo{author}{\bibfnamefont{S.~A.} \bibnamefont{Kivelson}},
  \bibnamefont{and} \bibinfo{author}{\bibfnamefont{S.~L.}
  \bibnamefont{Sondhi}}, \bibinfo{journal}{Phys. Rev. B}
  \textbf{\bibinfo{volume}{82}}, \bibinfo{pages}{035428}
  (\bibinfo{year}{2010}).

\bibitem[{\citenamefont{Lilly et~al.}(1999)\citenamefont{Lilly, Cooper,
  Eisenstein, Pfeiffer, and West}}]{Lilly-1999}
\bibinfo{author}{\bibfnamefont{M.~P.} \bibnamefont{Lilly}},
  \bibinfo{author}{\bibfnamefont{K.~B.} \bibnamefont{Cooper}},
  \bibinfo{author}{\bibfnamefont{J.~P.} \bibnamefont{Eisenstein}},
  \bibinfo{author}{\bibfnamefont{L.~N.} \bibnamefont{Pfeiffer}},
  \bibnamefont{and} \bibinfo{author}{\bibfnamefont{K.~W.} \bibnamefont{West}},
  \bibinfo{journal}{Phys. Rev. Lett.} \textbf{\bibinfo{volume}{82}},
  \bibinfo{pages}{394} (\bibinfo{year}{1999}).

\bibitem[{\citenamefont{Pan et~al.}(1999)\citenamefont{Pan, Du, Stormer, Tsui,
  Pfeiffer, Baldwin, and West}}]{Pan-1999}
\bibinfo{author}{\bibfnamefont{W.}~\bibnamefont{Pan}},
  \bibinfo{author}{\bibfnamefont{R.~R.} \bibnamefont{Du}},
  \bibinfo{author}{\bibfnamefont{H.~L.} \bibnamefont{Stormer}},
  \bibinfo{author}{\bibfnamefont{D.~C.} \bibnamefont{Tsui}},
  \bibinfo{author}{\bibfnamefont{L.~N.} \bibnamefont{Pfeiffer}},
  \bibinfo{author}{\bibfnamefont{K.~W.} \bibnamefont{Baldwin}},
  \bibnamefont{and} \bibinfo{author}{\bibfnamefont{K.~W.} \bibnamefont{West}},
  \bibinfo{journal}{Phys. Rev. Lett.} \textbf{\bibinfo{volume}{83}},
  \bibinfo{pages}{820} (\bibinfo{year}{1999}).

\bibitem[{\citenamefont{Cooper et~al.}(2002)\citenamefont{Cooper, Lilly,
  Eisenstein, Pfeiffer, and West}}]{Cooper-2002}
\bibinfo{author}{\bibfnamefont{K.~B.} \bibnamefont{Cooper}},
  \bibinfo{author}{\bibfnamefont{M.~P.} \bibnamefont{Lilly}},
  \bibinfo{author}{\bibfnamefont{J.~P.} \bibnamefont{Eisenstein}},
  \bibinfo{author}{\bibfnamefont{L.~N.} \bibnamefont{Pfeiffer}},
  \bibnamefont{and} \bibinfo{author}{\bibfnamefont{K.~W.} \bibnamefont{West}},
  \bibinfo{journal}{Phys. Rev. B} \textbf{\bibinfo{volume}{65}},
  \bibinfo{pages}{241313} (\bibinfo{year}{2002}).

\bibitem[{\citenamefont{Fradkin and Kivelson}(1999)}]{Fradkin-1999}
\bibinfo{author}{\bibfnamefont{E.}~\bibnamefont{Fradkin}} \bibnamefont{and}
  \bibinfo{author}{\bibfnamefont{S.}~\bibnamefont{Kivelson}},
  \bibinfo{journal}{Phys. Rev. B} \textbf{\bibinfo{volume}{59}},
  \bibinfo{pages}{8065} (\bibinfo{year}{1999}).

\bibitem[{\citenamefont{Fradkin et~al.}(2000)\citenamefont{Fradkin, Kivelson,
  Manousakis, and Nho}}]{Fradkin-2000}
\bibinfo{author}{\bibfnamefont{E.}~\bibnamefont{Fradkin}},
  \bibinfo{author}{\bibfnamefont{S.~A.} \bibnamefont{Kivelson}},
  \bibinfo{author}{\bibfnamefont{E.}~\bibnamefont{Manousakis}},
  \bibnamefont{and} \bibinfo{author}{\bibfnamefont{K.}~\bibnamefont{Nho}},
  \bibinfo{journal}{Phys. Rev. Lett.} \textbf{\bibinfo{volume}{84}},
  \bibinfo{pages}{1982} (\bibinfo{year}{2000}).

\bibitem[{\citenamefont{Xia et~al.}(2010)\citenamefont{Xia, Cvicek, Eisenstein,
  Pfeiffer, and West}}]{Xia-2010}
\bibinfo{author}{\bibfnamefont{J.}~\bibnamefont{Xia}},
  \bibinfo{author}{\bibfnamefont{V.}~\bibnamefont{Cvicek}},
  \bibinfo{author}{\bibfnamefont{J.~P.} \bibnamefont{Eisenstein}},
  \bibinfo{author}{\bibfnamefont{L.~N.} \bibnamefont{Pfeiffer}},
  \bibnamefont{and} \bibinfo{author}{\bibfnamefont{K.~W.} \bibnamefont{West}},
  \bibinfo{journal}{Phys. Rev. Lett.} \textbf{\bibinfo{volume}{105}},
  \bibinfo{pages}{176807} (\bibinfo{year}{2010}).

\bibitem[{\citenamefont{Xia et~al.}(2011)\citenamefont{Xia, Eisenstein,
  Pfeiffer, and West}}]{Xia-2011}
\bibinfo{author}{\bibfnamefont{J.}~\bibnamefont{Xia}},
  \bibinfo{author}{\bibfnamefont{J.~P.} \bibnamefont{Eisenstein}},
  \bibinfo{author}{\bibfnamefont{L.~N.} \bibnamefont{Pfeiffer}},
  \bibnamefont{and} \bibinfo{author}{\bibfnamefont{K.~W.} \bibnamefont{West}},
  \bibinfo{journal}{Nature Physics} \textbf{\bibinfo{volume}{7}},
  \bibinfo{pages}{845} (\bibinfo{year}{2011}).

\bibitem[{\citenamefont{Mulligan et~al.}(2010)\citenamefont{Mulligan, Nayak,
  and Kachru}}]{chetan}
\bibinfo{author}{\bibfnamefont{M.}~\bibnamefont{Mulligan}},
  \bibinfo{author}{\bibfnamefont{C.}~\bibnamefont{Nayak}}, \bibnamefont{and}
  \bibinfo{author}{\bibfnamefont{S.}~\bibnamefont{Kachru}},
  \bibinfo{journal}{Phys. Rev. B} \textbf{\bibinfo{volume}{82}},
  \bibinfo{pages}{085102} (\bibinfo{year}{2010}).

\bibitem[{\citenamefont{Mulligan et~al.}(2011)\citenamefont{Mulligan, Nayak,
  and Kachru}}]{Mulligan-2011}
\bibinfo{author}{\bibfnamefont{M.}~\bibnamefont{Mulligan}},
  \bibinfo{author}{\bibfnamefont{C.}~\bibnamefont{Nayak}}, \bibnamefont{and}
  \bibinfo{author}{\bibfnamefont{S.}~\bibnamefont{Kachru}},
  \bibinfo{journal}{Phys. Rev. B} \textbf{\bibinfo{volume}{84}},
  \bibinfo{pages}{195124} (\bibinfo{year}{2011}).

\bibitem[{\citenamefont{Balents}(1996)}]{Balents-1996}
\bibinfo{author}{\bibfnamefont{L.}~\bibnamefont{Balents}},
  \bibinfo{journal}{Europhysics Letters} \textbf{\bibinfo{volume}{33}},
  \bibinfo{pages}{291} (\bibinfo{year}{1996}).

\bibitem[{\citenamefont{Musaelian and Joynt}(1996)}]{Musaelian-1996}
\bibinfo{author}{\bibfnamefont{K.}~\bibnamefont{Musaelian}} \bibnamefont{and}
  \bibinfo{author}{\bibfnamefont{R.}~\bibnamefont{Joynt}},
  \bibinfo{journal}{Journal of Physics: Condensed Matter}
  \textbf{\bibinfo{volume}{8}}, \bibinfo{pages}{L105} (\bibinfo{year}{1996}).

\bibitem[{\citenamefont{Haldane}(2011)}]{Haldane-2011}
\bibinfo{author}{\bibfnamefont{F.~D.~M.} \bibnamefont{Haldane}},
  \bibinfo{journal}{Phys. Rev. Lett.} \textbf{\bibinfo{volume}{107}},
  \bibinfo{pages}{116801} (\bibinfo{year}{2011}).

\bibitem[{\citenamefont{Yang et~al.}(2012)\citenamefont{Yang, Papi{\'c},
  Rezayi, Bhatt, and Haldane}}]{haldanemetric}
\bibinfo{author}{\bibfnamefont{B.}~\bibnamefont{Yang}},
  \bibinfo{author}{\bibfnamefont{Z.}~\bibnamefont{Papi{\'c}}},
  \bibinfo{author}{\bibfnamefont{E.~H.} \bibnamefont{Rezayi}},
  \bibinfo{author}{\bibfnamefont{R.~N.} \bibnamefont{Bhatt}}, \bibnamefont{and}
  \bibinfo{author}{\bibfnamefont{F.~D.~M.} \bibnamefont{Haldane}},
  \bibinfo{journal}{Phys. Rev. B} \textbf{\bibinfo{volume}{85}},
  \bibinfo{pages}{165318} (\bibinfo{year}{2012}).

\bibitem[{\citenamefont{Qiu et~al.}(2012)\citenamefont{Qiu, Haldane, Wan, Yang,
  and Yi}}]{Qiu-2012}
\bibinfo{author}{\bibfnamefont{R.-Z.} \bibnamefont{Qiu}},
  \bibinfo{author}{\bibfnamefont{F.~D.~M.} \bibnamefont{Haldane}},
  \bibinfo{author}{\bibfnamefont{X.}~\bibnamefont{Wan}},
  \bibinfo{author}{\bibfnamefont{K.}~\bibnamefont{Yang}}, \bibnamefont{and}
  \bibinfo{author}{\bibfnamefont{S.}~\bibnamefont{Yi}}, \bibinfo{journal}{Phys.
  Rev. B} \textbf{\bibinfo{volume}{85}}, \bibinfo{pages}{115308}
  (\bibinfo{year}{2012}).

\bibitem[{\citenamefont{Maciejko et~al.}(2013)\citenamefont{Maciejko, Hsu,
  Kivelson, Park, and Sondhi}}]{Maciejko-2013}
\bibinfo{author}{\bibfnamefont{J.}~\bibnamefont{Maciejko}},
  \bibinfo{author}{\bibfnamefont{B.}~\bibnamefont{Hsu}},
  \bibinfo{author}{\bibfnamefont{S.~A.} \bibnamefont{Kivelson}},
  \bibinfo{author}{\bibfnamefont{Y.}~\bibnamefont{Park}}, \bibnamefont{and}
  \bibinfo{author}{\bibfnamefont{S.~L.} \bibnamefont{Sondhi}},
  \bibinfo{journal}{Phys. Rev. B} \textbf{\bibinfo{volume}{88}},
  \bibinfo{pages}{125137} (\bibinfo{year}{2013}).

\bibitem[{\citenamefont{Barci and Fradkin}(2011)}]{Barci-2011}
\bibinfo{author}{\bibfnamefont{D.~G.} \bibnamefont{Barci}} \bibnamefont{and}
  \bibinfo{author}{\bibfnamefont{E.}~\bibnamefont{Fradkin}},
  \bibinfo{journal}{Phys. Rev. B} \textbf{\bibinfo{volume}{83}},
  \bibinfo{pages}{100509} (\bibinfo{year}{2011}).

\bibitem[{\citenamefont{Avron et~al.}(1995)\citenamefont{Avron, Seiler, and
  Zograf}}]{avron1995}
\bibinfo{author}{\bibfnamefont{J.~E.} \bibnamefont{Avron}},
  \bibinfo{author}{\bibfnamefont{R.}~\bibnamefont{Seiler}}, \bibnamefont{and}
  \bibinfo{author}{\bibfnamefont{P.~G.} \bibnamefont{Zograf}},
  \bibinfo{journal}{Phys. Rev. Lett.} \textbf{\bibinfo{volume}{75}},
  \bibinfo{pages}{697} (\bibinfo{year}{1995}).

\bibitem[{\citenamefont{Read}(2009)}]{Read-2009}
\bibinfo{author}{\bibfnamefont{N.}~\bibnamefont{Read}}, \bibinfo{journal}{Phys.
  Rev B} \textbf{\bibinfo{volume}{79}}, \bibinfo{pages}{045308}
  (\bibinfo{year}{2009}).

\bibitem[{\citenamefont{Hughes et~al.}(2011)\citenamefont{Hughes, Leigh, and
  Fradkin}}]{taylor}
\bibinfo{author}{\bibfnamefont{T.~L.} \bibnamefont{Hughes}},
  \bibinfo{author}{\bibfnamefont{R.~G.} \bibnamefont{Leigh}}, \bibnamefont{and}
  \bibinfo{author}{\bibfnamefont{E.}~\bibnamefont{Fradkin}},
  \bibinfo{journal}{Phys. Rev. Lett.} \textbf{\bibinfo{volume}{107}},
  \bibinfo{pages}{075502} (\bibinfo{year}{2011}).

\bibitem[{\citenamefont{Nicolis and Son}(2011)}]{Nicolis-2011}
\bibinfo{author}{\bibfnamefont{A.}~\bibnamefont{Nicolis}} \bibnamefont{and}
  \bibinfo{author}{\bibfnamefont{D.~T.} \bibnamefont{Son}},
  \emph{\bibinfo{title}{{Hall viscosity from effective field theory}}}
  (\bibinfo{year}{2011}), \bibinfo{note}{unpublished},
  \eprint{arXiv:1103.2137}.

\bibitem[{\citenamefont{Hoyos and Son}(2012)}]{vis}
\bibinfo{author}{\bibfnamefont{C.}~\bibnamefont{Hoyos}} \bibnamefont{and}
  \bibinfo{author}{\bibfnamefont{D.~T.} \bibnamefont{Son}},
  \bibinfo{journal}{Phys. Rev. Lett.} \textbf{\bibinfo{volume}{108}},
  \bibinfo{pages}{066805} (\bibinfo{year}{2012}).

\bibitem[{\citenamefont{Sun et~al.}(2009)\citenamefont{Sun, Yao, Fradkin, and
  Kivelson}}]{kai}
\bibinfo{author}{\bibfnamefont{K.}~\bibnamefont{Sun}},
  \bibinfo{author}{\bibfnamefont{H.}~\bibnamefont{Yao}},
  \bibinfo{author}{\bibfnamefont{E.}~\bibnamefont{Fradkin}}, \bibnamefont{and}
  \bibinfo{author}{\bibfnamefont{S.~A.} \bibnamefont{Kivelson}},
  \bibinfo{journal}{Phys. Rev. Lett.} \textbf{\bibinfo{volume}{103}},
  \bibinfo{pages}{046811} (\bibinfo{year}{2009}).

\bibitem[{\citenamefont{You and Fradkin}(2013)}]{You-2013}
\bibinfo{author}{\bibfnamefont{Y.}~\bibnamefont{You}} \bibnamefont{and}
  \bibinfo{author}{\bibfnamefont{E.}~\bibnamefont{Fradkin}},
  \emph{\bibinfo{title}{{The Nematic Fractional Quantum Hall State}}}
  (\bibinfo{year}{2013}), \bibinfo{note}{in preparation}.

\bibitem[{\citenamefont{Ardonne et~al.}(2004)\citenamefont{Ardonne, Fendley,
  and Fradkin}}]{Ardonne2004}
\bibinfo{author}{\bibfnamefont{E.}~\bibnamefont{Ardonne}},
  \bibinfo{author}{\bibfnamefont{P.}~\bibnamefont{Fendley}}, \bibnamefont{and}
  \bibinfo{author}{\bibfnamefont{E.}~\bibnamefont{Fradkin}},
  \bibinfo{journal}{Annals of Physics} \textbf{\bibinfo{volume}{310}},
  \bibinfo{pages}{493} (\bibinfo{year}{2004}).

\bibitem[{\citenamefont{Raghu et~al.}(2008)\citenamefont{Raghu, Qi, Honerkamp,
  and Zhang}}]{qi}
\bibinfo{author}{\bibfnamefont{S.}~\bibnamefont{Raghu}},
  \bibinfo{author}{\bibfnamefont{X.-L.} \bibnamefont{Qi}},
  \bibinfo{author}{\bibfnamefont{C.}~\bibnamefont{Honerkamp}},
  \bibnamefont{and} \bibinfo{author}{\bibfnamefont{S.-C.} \bibnamefont{Zhang}},
  \bibinfo{journal}{Phys. Rev. Lett.} \textbf{\bibinfo{volume}{100}},
  \bibinfo{pages}{156401} (\bibinfo{year}{2008}).

\bibitem[{\citenamefont{Nandkishore and Levitov}(2010)}]{Nandkishore2010}
\bibinfo{author}{\bibfnamefont{R.}~\bibnamefont{Nandkishore}} \bibnamefont{and}
  \bibinfo{author}{\bibfnamefont{L.}~\bibnamefont{Levitov}},
  \bibinfo{journal}{Phys. Rev. B} \textbf{\bibinfo{volume}{82}},
  \bibinfo{pages}{115124} (\bibinfo{year}{2010}).

\bibitem[{\citenamefont{Vafek and Yang}(2010)}]{Vafek2010}
\bibinfo{author}{\bibfnamefont{O.}~\bibnamefont{Vafek}} \bibnamefont{and}
  \bibinfo{author}{\bibfnamefont{K.}~\bibnamefont{Yang}},
  \bibinfo{journal}{Phys. Rev. B} \textbf{\bibinfo{volume}{81}},
  \bibinfo{pages}{041401(R)} (\bibinfo{year}{2010}).

\bibitem[{\citenamefont{Lemonik et~al.}(2012)\citenamefont{Lemonik, Aleiner,
  and Fal'ko}}]{Lemonik2012}
\bibinfo{author}{\bibfnamefont{Y.}~\bibnamefont{Lemonik}},
  \bibinfo{author}{\bibfnamefont{I.}~\bibnamefont{Aleiner}}, \bibnamefont{and}
  \bibinfo{author}{\bibfnamefont{V.~I.} \bibnamefont{Fal'ko}},
  \bibinfo{journal}{Phys. Rev. B} \textbf{\bibinfo{volume}{85}},
  \bibinfo{pages}{245451} (\bibinfo{year}{2012}).

\bibitem[{\citenamefont{Fu}(2011)}]{Fu2011}
\bibinfo{author}{\bibfnamefont{L.}~\bibnamefont{Fu}}, \bibinfo{journal}{Phys.
  Rev. Lett.} \textbf{\bibinfo{volume}{106}}, \bibinfo{pages}{106802}
  (\bibinfo{year}{2011}).

\bibitem[{\citenamefont{Fang et~al.}(2012)\citenamefont{Fang, Gilbert, Su,
  Bernevig, and Hasan}}]{Fang2012}
\bibinfo{author}{\bibfnamefont{C.}~\bibnamefont{Fang}},
  \bibinfo{author}{\bibfnamefont{M.~J.} \bibnamefont{Gilbert}},
  \bibinfo{author}{\bibfnamefont{S.~Y.} \bibnamefont{Su}},
  \bibinfo{author}{\bibfnamefont{B.~A.} \bibnamefont{Bernevig}},
  \bibnamefont{and} \bibinfo{author}{\bibfnamefont{M.~Z.} \bibnamefont{Hasan}},
  \emph{\bibinfo{title}{{Surface State Quasiparticle Interference in
  Crystalline Topological Insulators}}} (\bibinfo{year}{2012}),
  \bibinfo{note}{unpublished}, \eprint{arXiv:1212.3285}.

\bibitem[{\citenamefont{Sun et~al.}(2011)\citenamefont{Sun, Gu, Katsura, and
  {Das Sarma}}}]{sun2011}
\bibinfo{author}{\bibfnamefont{K.}~\bibnamefont{Sun}},
  \bibinfo{author}{\bibfnamefont{Z.}~\bibnamefont{Gu}},
  \bibinfo{author}{\bibfnamefont{H.}~\bibnamefont{Katsura}}, \bibnamefont{and}
  \bibinfo{author}{\bibfnamefont{S.}~\bibnamefont{{Das Sarma}}},
  \bibinfo{journal}{Phys. Rev. Lett.} \textbf{\bibinfo{volume}{106}},
  \bibinfo{pages}{236803} (\bibinfo{year}{2011}).

\bibitem[{\citenamefont{Tsai et~al.}(2011)\citenamefont{Tsai, Fang, Yao, and
  Hu}}]{tsai-2011}
\bibinfo{author}{\bibfnamefont{W.-F.} \bibnamefont{Tsai}},
  \bibinfo{author}{\bibfnamefont{C.}~\bibnamefont{Fang}},
  \bibinfo{author}{\bibfnamefont{H.}~\bibnamefont{Yao}}, \bibnamefont{and}
  \bibinfo{author}{\bibfnamefont{J.}~\bibnamefont{Hu}},
  \emph{\bibinfo{title}{{Interaction-driven topological and nematic phases on
  the Lieb lattice}}} (\bibinfo{year}{2011}), \bibinfo{note}{unpublished},
  \eprint{arXiv:1112.5789}.

\bibitem[{\citenamefont{Vafek}(2010)}]{Vafek2010b}
\bibinfo{author}{\bibfnamefont{O.}~\bibnamefont{Vafek}},
  \bibinfo{journal}{Phys. Rev. B} \textbf{\bibinfo{volume}{82}},
  \bibinfo{pages}{205106} (\bibinfo{year}{2010}).

\bibitem[{\citenamefont{Wilson}(1970)}]{wilson-1970}
\bibinfo{author}{\bibfnamefont{K.~G.} \bibnamefont{Wilson}},
  \bibinfo{journal}{Phys. Rev. D} \textbf{\bibinfo{volume}{2}},
  \bibinfo{pages}{1478} (\bibinfo{year}{1970}).

\bibitem[{\citenamefont{Abrikosov}(1974)}]{Abrikosov1974}
\bibinfo{author}{\bibfnamefont{A.~A.} \bibnamefont{Abrikosov}},
  \bibinfo{journal}{Sov. Phys. JETP} \textbf{\bibinfo{volume}{39}},
  \bibinfo{pages}{709} (\bibinfo{year}{1974}).

\bibitem[{\citenamefont{Xu et~al.}(2012)\citenamefont{Xu, Liu, Alidoust,
  Neupane, Qian, Belopolski, Denlinger, Wang, Lin, Wray et~al.}}]{Hasan2012}
\bibinfo{author}{\bibfnamefont{S.-Y.} \bibnamefont{Xu}},
  \bibinfo{author}{\bibfnamefont{C.}~\bibnamefont{Liu}},
  \bibinfo{author}{\bibfnamefont{N.}~\bibnamefont{Alidoust}},
  \bibinfo{author}{\bibfnamefont{M.}~\bibnamefont{Neupane}},
  \bibinfo{author}{\bibfnamefont{D.}~\bibnamefont{Qian}},
  \bibinfo{author}{\bibnamefont{Belopolski}},
  \bibinfo{author}{\bibfnamefont{J.~D.} \bibnamefont{Denlinger}},
  \bibinfo{author}{\bibfnamefont{Y.~J.} \bibnamefont{Wang}},
  \bibinfo{author}{\bibfnamefont{H.}~\bibnamefont{Lin}},
  \bibinfo{author}{\bibfnamefont{L.~A.} \bibnamefont{Wray}},
  \bibnamefont{et~al.}, \bibinfo{journal}{Nature Communications}
  \textbf{\bibinfo{volume}{3}}, \bibinfo{pages}{1192} (\bibinfo{year}{2012}).

\bibitem[{\citenamefont{Gyenis et~al.}(2013)\citenamefont{Gyenis, Drozdov,
  Nadj-Perge, Jeong, Seo, Pletikosic, Valla, Gu, and Yazdani}}]{Yazdani2013}
\bibinfo{author}{\bibfnamefont{A.}~\bibnamefont{Gyenis}},
  \bibinfo{author}{\bibfnamefont{I.~K.} \bibnamefont{Drozdov}},
  \bibinfo{author}{\bibfnamefont{S.}~\bibnamefont{Nadj-Perge}},
  \bibinfo{author}{\bibfnamefont{O.~B.} \bibnamefont{Jeong}},
  \bibinfo{author}{\bibfnamefont{J.}~\bibnamefont{Seo}},
  \bibinfo{author}{\bibfnamefont{I.}~\bibnamefont{Pletikosic}},
  \bibinfo{author}{\bibfnamefont{T.}~\bibnamefont{Valla}},
  \bibinfo{author}{\bibfnamefont{G.~D.} \bibnamefont{Gu}}, \bibnamefont{and}
  \bibinfo{author}{\bibfnamefont{A.}~\bibnamefont{Yazdani}},
  \emph{\bibinfo{title}{{Quasiparticle Interference on the Surface of
  Topological Crystalline Insulator Pb$_{1-x}$Sn$_x$Se}}}
  (\bibinfo{year}{2013}), \bibinfo{note}{unpublished},
  \eprint{arXiv:1306.0043}.

\bibitem[{\citenamefont{Yang and Kim}(2010)}]{Yang2010}
\bibinfo{author}{\bibfnamefont{B.-J.} \bibnamefont{Yang}} \bibnamefont{and}
  \bibinfo{author}{\bibfnamefont{Y.~B.} \bibnamefont{Kim}},
  \bibinfo{journal}{Phys. Rev. B} \textbf{\bibinfo{volume}{82}},
  \bibinfo{pages}{085111} (\bibinfo{year}{2010}).

\bibitem[{\citenamefont{Witczak-Krempa and Kim}(2012)}]{Witczak-Krempa2012}
\bibinfo{author}{\bibfnamefont{W.}~\bibnamefont{Witczak-Krempa}}
  \bibnamefont{and} \bibinfo{author}{\bibfnamefont{Y.~B.} \bibnamefont{Kim}},
  \bibinfo{journal}{Phys. Rev. B} \textbf{\bibinfo{volume}{85}},
  \bibinfo{pages}{045124} (\bibinfo{year}{2012}).

\bibitem[{\citenamefont{Witczak-Krempa
  et~al.}(2013)\citenamefont{Witczak-Krempa, Chen, Kim, and
  Balents}}]{Witczak-Krempa2013}
\bibinfo{author}{\bibfnamefont{W.}~\bibnamefont{Witczak-Krempa}},
  \bibinfo{author}{\bibfnamefont{G.}~\bibnamefont{Chen}},
  \bibinfo{author}{\bibfnamefont{Y.~B.} \bibnamefont{Kim}}, \bibnamefont{and}
  \bibinfo{author}{\bibfnamefont{L.}~\bibnamefont{Balents}},
  \emph{\bibinfo{title}{Correlated quantum phenomena in the strong spin-orbit
  regime}} (\bibinfo{year}{2013}), \bibinfo{note}{unpublished},
  \eprint{arXiv:1305.2193}.

\bibitem[{\citenamefont{Chan et~al.}(2013)\citenamefont{Chan, Hughes, Ryu, and
  Fradkin}}]{atma}
\bibinfo{author}{\bibfnamefont{A.}~\bibnamefont{Chan}},
  \bibinfo{author}{\bibfnamefont{T.~L.} \bibnamefont{Hughes}},
  \bibinfo{author}{\bibfnamefont{S.}~\bibnamefont{Ryu}}, \bibnamefont{and}
  \bibinfo{author}{\bibfnamefont{E.}~\bibnamefont{Fradkin}},
  \bibinfo{journal}{Phys. Rev. B} \textbf{\bibinfo{volume}{87}},
  \bibinfo{pages}{085132} (\bibinfo{year}{2013}).

\bibitem[{\citenamefont{Fradkin and Schaposnik}(1994)}]{Fradkin1994}
\bibinfo{author}{\bibfnamefont{E.}~\bibnamefont{Fradkin}} \bibnamefont{and}
  \bibinfo{author}{\bibfnamefont{F.~A.} \bibnamefont{Schaposnik}},
  \bibinfo{journal}{Phys. Lett. B} \textbf{\bibinfo{volume}{338}},
  \bibinfo{pages}{253} (\bibinfo{year}{1994}).

\bibitem[{\citenamefont{{Le Guillou} et~al.}(1997)\citenamefont{{Le Guillou},
  Moreno, Schaposnik, and N{\'u}{\~n}ez}}]{LeGuillou1997}
\bibinfo{author}{\bibfnamefont{J.~C.} \bibnamefont{{Le Guillou}}},
  \bibinfo{author}{\bibfnamefont{E.}~\bibnamefont{Moreno}},
  \bibinfo{author}{\bibfnamefont{F.~A.} \bibnamefont{Schaposnik}},
  \bibnamefont{and}
  \bibinfo{author}{\bibfnamefont{C.}~\bibnamefont{N{\'u}{\~n}ez}},
  \bibinfo{journal}{Phys. Lett. B} \textbf{\bibinfo{volume}{409}},
  \bibinfo{pages}{257} (\bibinfo{year}{1997}).

\bibitem[{\citenamefont{Burgess et~al.}(1994)\citenamefont{Burgess, L{\"u}tken,
  and Quevedo}}]{Burgess1994}
\bibinfo{author}{\bibfnamefont{C.~P.} \bibnamefont{Burgess}},
  \bibinfo{author}{\bibfnamefont{C.~A.} \bibnamefont{L{\"u}tken}},
  \bibnamefont{and} \bibinfo{author}{\bibfnamefont{F.}~\bibnamefont{Quevedo}},
  \bibinfo{journal}{Physics Letters B} \textbf{\bibinfo{volume}{336}},
  \bibinfo{pages}{18} (\bibinfo{year}{1994}).

\bibitem[{\citenamefont{Cho and Moore}(2011)}]{Cho2011}
\bibinfo{author}{\bibfnamefont{G.~Y.} \bibnamefont{Cho}} \bibnamefont{and}
  \bibinfo{author}{\bibfnamefont{J.~E.} \bibnamefont{Moore}},
  \bibinfo{journal}{Ann. Phys.} \textbf{\bibinfo{volume}{326}},
  \bibinfo{pages}{1515} (\bibinfo{year}{2011}).

\bibitem[{\citenamefont{Fr{\"o}hlich and Zee}(1991)}]{Frohlich1991b}
\bibinfo{author}{\bibfnamefont{J.}~\bibnamefont{Fr{\"o}hlich}}
  \bibnamefont{and} \bibinfo{author}{\bibfnamefont{A.}~\bibnamefont{Zee}},
  \bibinfo{journal}{Nucl. Phys. B} \textbf{\bibinfo{volume}{364}},
  \bibinfo{pages}{517} (\bibinfo{year}{1991}).

\bibitem[{\citenamefont{Wen and Zee}(1992{\natexlab{b}})}]{Wen1992}
\bibinfo{author}{\bibfnamefont{X.-G.} \bibnamefont{Wen}} \bibnamefont{and}
  \bibinfo{author}{\bibfnamefont{A.}~\bibnamefont{Zee}},
  \bibinfo{journal}{Phys. Rev. B} \textbf{\bibinfo{volume}{46}},
  \bibinfo{pages}{2290} (\bibinfo{year}{1992}{\natexlab{b}}).

\bibitem[{\citenamefont{Wen}(1995{\natexlab{b}})}]{Wen1995}
\bibinfo{author}{\bibfnamefont{X.~G.} \bibnamefont{Wen}},
  \bibinfo{journal}{Adv. Phys.} \textbf{\bibinfo{volume}{44}},
  \bibinfo{pages}{405} (\bibinfo{year}{1995}{\natexlab{b}}).

\bibitem[{\citenamefont{Moshe and Zinn-Justin}(2003)}]{Moshe2003}
\bibinfo{author}{\bibfnamefont{M.}~\bibnamefont{Moshe}} \bibnamefont{and}
  \bibinfo{author}{\bibfnamefont{J.}~\bibnamefont{Zinn-Justin}},
  \bibinfo{journal}{Phys. Rept.} \textbf{\bibinfo{volume}{385}},
  \bibinfo{pages}{69} (\bibinfo{year}{2003}).

\bibitem[{\citenamefont{Oganesyan et~al.}(2001)\citenamefont{Oganesyan,
  Kivelson, and Fradkin}}]{Oganesyan-2001}
\bibinfo{author}{\bibfnamefont{V.}~\bibnamefont{Oganesyan}},
  \bibinfo{author}{\bibfnamefont{S.~A.} \bibnamefont{Kivelson}},
  \bibnamefont{and} \bibinfo{author}{\bibfnamefont{E.}~\bibnamefont{Fradkin}},
  \bibinfo{journal}{Phys. Rev. B} \textbf{\bibinfo{volume}{64}},
  \bibinfo{pages}{195109} (\bibinfo{year}{2001}).

\bibitem[{\citenamefont{Chaikin and Lubensky}(1995)}]{chaikin-1995}
\bibinfo{author}{\bibfnamefont{P.~M.} \bibnamefont{Chaikin}} \bibnamefont{and}
  \bibinfo{author}{\bibfnamefont{T.~C.} \bibnamefont{Lubensky}},
  \emph{\bibinfo{title}{{Principles of Condensed Matter Physics}}}
  (\bibinfo{publisher}{Cambridge University Press},
  \bibinfo{address}{Cambridge, UK}, \bibinfo{year}{1995}).

\bibitem[{\citenamefont{Deser et~al.}(1982)\citenamefont{Deser, Jackiw, and
  Templeton}}]{Deser-1982}
\bibinfo{author}{\bibfnamefont{S.}~\bibnamefont{Deser}},
  \bibinfo{author}{\bibfnamefont{R.}~\bibnamefont{Jackiw}}, \bibnamefont{and}
  \bibinfo{author}{\bibfnamefont{S.}~\bibnamefont{Templeton}},
  \bibinfo{journal}{Phys. Rev. Lett.} \textbf{\bibinfo{volume}{48}},
  \bibinfo{pages}{975} (\bibinfo{year}{1982}).

\bibitem[{\citenamefont{Redlich}(1984)}]{Redlich-1984}
\bibinfo{author}{\bibfnamefont{A.~N.} \bibnamefont{Redlich}},
  \bibinfo{journal}{Phys. Rev. D} \textbf{\bibinfo{volume}{29}},
  \bibinfo{pages}{2366} (\bibinfo{year}{1984}).

\bibitem[{\citenamefont{de~Gennes and Prost}(1993)}]{deGennes-1993}
\bibinfo{author}{\bibfnamefont{P.~G.} \bibnamefont{de~Gennes}}
  \bibnamefont{and} \bibinfo{author}{\bibfnamefont{J.}~\bibnamefont{Prost}},
  \emph{\bibinfo{title}{{The Physics of Liquid Crystals}}}
  (\bibinfo{publisher}{Oxford Sci./Clarendon}, \bibinfo{address}{Oxford, UK},
  \bibinfo{year}{1993}).

\bibitem[{\citenamefont{Hughes et~al.}(2013)\citenamefont{Hughes, Leigh, and
  Parrikar}}]{Hughes-2013}
\bibinfo{author}{\bibfnamefont{T.~L.} \bibnamefont{Hughes}},
  \bibinfo{author}{\bibfnamefont{R.~G.} \bibnamefont{Leigh}}, \bibnamefont{and}
  \bibinfo{author}{\bibfnamefont{O.}~\bibnamefont{Parrikar}},
  \bibinfo{journal}{Phys. Rev. D} \textbf{\bibinfo{volume}{88}},
  \bibinfo{pages}{025040} (\bibinfo{year}{2013}).

\bibitem[{\citenamefont{Hsieh et~al.}(2012)\citenamefont{Hsieh, L., Liu, Duan,
  Bansil, and Fu}}]{Hsieh2012}
\bibinfo{author}{\bibfnamefont{T.~H.} \bibnamefont{Hsieh}},
  \bibinfo{author}{\bibfnamefont{H.}~\bibnamefont{L.}},
  \bibinfo{author}{\bibfnamefont{J.~W.} \bibnamefont{Liu}},
  \bibinfo{author}{\bibfnamefont{W.~H.} \bibnamefont{Duan}},
  \bibinfo{author}{\bibfnamefont{A.}~\bibnamefont{Bansil}}, \bibnamefont{and}
  \bibinfo{author}{\bibfnamefont{L.}~\bibnamefont{Fu}},
  \bibinfo{journal}{Nature Communications} \textbf{\bibinfo{volume}{3}},
  \bibinfo{pages}{982} (\bibinfo{year}{2012}).

\bibitem[{\citenamefont{Dziawa et~al.}(2012)\citenamefont{Dziawa, Kowalski,
  Dybko, Buczko, Szczerbakow, Szot, Lusakowska, Balasubramanian, Wojek,
  Berntsen et~al.}}]{Story2012}
\bibinfo{author}{\bibfnamefont{P.}~\bibnamefont{Dziawa}},
  \bibinfo{author}{\bibfnamefont{B.~J.} \bibnamefont{Kowalski}},
  \bibinfo{author}{\bibfnamefont{K.}~\bibnamefont{Dybko}},
  \bibinfo{author}{\bibfnamefont{R.}~\bibnamefont{Buczko}},
  \bibinfo{author}{\bibfnamefont{A.}~\bibnamefont{Szczerbakow}},
  \bibinfo{author}{\bibfnamefont{M.}~\bibnamefont{Szot}},
  \bibinfo{author}{\bibfnamefont{E.}~\bibnamefont{Lusakowska}},
  \bibinfo{author}{\bibfnamefont{T.}~\bibnamefont{Balasubramanian}},
  \bibinfo{author}{\bibfnamefont{B.~M.} \bibnamefont{Wojek}},
  \bibinfo{author}{\bibfnamefont{M.~H.} \bibnamefont{Berntsen}},
  \bibnamefont{et~al.}, \bibinfo{journal}{Nature Materials}
  \textbf{\bibinfo{volume}{11}}, \bibinfo{pages}{1023} (\bibinfo{year}{2012}).

\bibitem[{\citenamefont{Okada et~al.}(2013)\citenamefont{Okada, Serbyn, Lin,
  Walkup, Zhou, Dhital, Neupane, Xu, Wang, Sankar et~al.}}]{Hasan2013}
\bibinfo{author}{\bibfnamefont{Y.}~\bibnamefont{Okada}},
  \bibinfo{author}{\bibfnamefont{M.}~\bibnamefont{Serbyn}},
  \bibinfo{author}{\bibfnamefont{H.}~\bibnamefont{Lin}},
  \bibinfo{author}{\bibfnamefont{D.}~\bibnamefont{Walkup}},
  \bibinfo{author}{\bibfnamefont{W.~W.} \bibnamefont{Zhou}},
  \bibinfo{author}{\bibfnamefont{C.}~\bibnamefont{Dhital}},
  \bibinfo{author}{\bibfnamefont{M.}~\bibnamefont{Neupane}},
  \bibinfo{author}{\bibfnamefont{S.~Y.} \bibnamefont{Xu}},
  \bibinfo{author}{\bibfnamefont{Y.~J.} \bibnamefont{Wang}},
  \bibinfo{author}{\bibfnamefont{R.}~\bibnamefont{Sankar}},
  \bibnamefont{et~al.}, \emph{\bibinfo{title}{{Observation of Dirac node
  formation and mass acquisition in a topological crystalline insulator}}}
  (\bibinfo{year}{2013}), \eprint{arXiv:1305.2823}.

\bibitem[{\citenamefont{Wan et~al.}(2011)\citenamefont{Wan, Turner, Vishwanath,
  and Savrasov}}]{Wan-2011}
\bibinfo{author}{\bibfnamefont{X.}~\bibnamefont{Wan}},
  \bibinfo{author}{\bibfnamefont{A.~M.} \bibnamefont{Turner}},
  \bibinfo{author}{\bibfnamefont{A.}~\bibnamefont{Vishwanath}},
  \bibnamefont{and} \bibinfo{author}{\bibfnamefont{S.~Y.}
  \bibnamefont{Savrasov}}, \bibinfo{journal}{Phys. Rev. B}
  \textbf{\bibinfo{volume}{83}}, \bibinfo{pages}{205101}
  (\bibinfo{year}{2011}).

\end{thebibliography}

\end{document}